\documentclass[epj,nopacs]{svjour}
\usepackage{microtype,ragged2e}
\usepackage{amsfonts,amsmath,amssymb,bbold,mathtools}
\newcommand{\maths}[1]{$\smash{#1}$}
\usepackage{booktabs,cite,cuted,graphicx,xcolor}
\usepackage[pdfstartview=FitH,colorlinks=true,citecolor=blue,linkcolor=blue,urlcolor=blue]{hyperref}
\allowdisplaybreaks[4]

\begin{document}

\sloppypar

\title{Critical speed of a binary superfluid of light}

\author{
Pierre-\'Elie Larr\'e
\inst{1}
\thanks{\email{pierre-elie.larre@universite-paris-saclay.fr} (corresponding author)}
\and
Claire Michel
\inst{2,3}
\thanks{\email{claire.michel@univ-cotedazur.fr}}
\and
Nicolas Cherroret
\inst{4}
\thanks{\email{nicolas.cherroret@lkb.upmc.fr}}
}

\institute{
Universit\'e Paris-Saclay, CNRS, LPTMS, 91405, Orsay, France
\and
Universit\'e C\^ote d'Azur, CNRS, INPHYNI, Nice, France
\and
Institut Universitaire de France (IUF)
\and
Laboratoire Kastler Brossel, Sorbonne Universit\'e, CNRS, ENS-PSL Research University, Coll\`ege de France; 4 Place Jussieu, 75005 Paris, France
}

\date{\today}

\abstract{
We theoretically study the critical speed for superfluid flow of a two-dimensional miscible binary superfluid of light past a polarization-sensitive optical obstacle. This speed corresponds to the maximum mean flow velocity below which dissipation is absent. In the weak-obstacle regime, linear-response theory shows that the critical speed is set by Landau's criterion applied to the density and spin Bogoliubov modes, whose relative ordering can be inverted due to saturation of the optical nonlinearity. For obstacles of arbitrary strength and large spatial extent, we determine the critical speed from the conditions for strong ellipticity of the stationary hydrodynamic equations within the hydraulic and incompressible approximations. Numerical simulations in this regime reveal that the breakdown of superfluidity is initiated by the nucleation of vortex-antivortex pairs for an impenetrable obstacle, and of Jones-Roberts solitons for a penetrable obstacle. Beyond superfluids of light, our results provide a general framework for the critical speed of two-dimensional binary nonlinear Schr\"odinger superflows, including Bose-Bose quantum mixtures.
}

\maketitle

\section{Introduction}
\label{Sec:Intro}

Realized in liquid helium~\cite{Leggett1999}, quantum gases~\cite{Pitaevskii2016}, and polariton condensates~\cite{Carusotto2013}, superfluids provide an ideal platform for exploring macroscopic quantum phenomena, ranging from dissipationless flow to collective and topological excitations. Among these systems, \textit{Bose-Bose superfluid mixtures} have attracted considerable research interest. This is reflected in a wide range of experimental investigations on quantum hydrodynamics~\cite{Fava2018, Wilson2021, Cavicchioli2022}, sound modes~\cite{Kim2020, Kim2021, Cominotti2022}, topological defects~\cite{Farolfi2020, Bresolin2023, Mossman2024}, and miscibility problems~\cite{Hall1998, Miesner1999, Papp2008, Burchianti2020} (see also Refs.~\cite{Pitaevskii2016, Marti2016, Recati2016, Recati2022, Baroni2024, QuantumMixtures2025} and~\cite{Pu1998, Sinatra2000, Bouchoule2003, Palencia2013, Hofmann2014, Mathey2023, Gliott2025} for comprehensive reviews and theoretical background).

These binary superfluids typically consist of two distinct internal states (or, in some cases, two different species) whose coupled coherent dynamics can be described by two macroscopic wave functions, \maths{\psi_{+}} and \maths{\psi_{-}}. The squared modulus and phase gradient of \maths{\psi_{\pm}} yield the hydrodynamic density \maths{\rho_{\pm}} and velocity \maths{\mathbf{v}_{\pm}}, respectively, of the components in the \maths{\pm} state. This dual composition enriches the system's overall behavior, giving rise to two distinct yet typically hybridized excitation modes: a \textit{density mode}, involving fluctuations in the superfluid's total density \maths{\rho_{+}+\rho_{-}} and velocity \maths{\mathbf{v}_{+}+\mathbf{v}_{-}}; and a \textit{spin mode}, associated with variations in the relative density \maths{\rho_{+}-\rho_{-}} and velocity \maths{\mathbf{v}_{+}-\mathbf{v}_{-}}~\cite{Pitaevskii2016}.

Intriguingly, the physics governing these binary quantum mixtures finds a remarkable counterpart in a seemingly unrelated classical system: the paraxial propagation of a two-polarization laser beam in a nonlinear medium~\cite{Boyd2020, Ackemann1999, Agrawal2019, Barad1997}. In this all-optical platform, the two orthogonal polarization states of light behave analogously to two distinct superfluid components, exhibiting a coupled coherent dynamics akin to that of Bose-Bose superfluid mixtures. This analogy extends to the polarization degrees of freedom of light the well-known resemblance between the nonlinear Schr\"odinger equation of scalar paraxial optics~\cite{Boyd2020} and the Gross-Pitaevskii equation for scalar Bose-Einstein condensates~\cite{Pitaevskii2016}, a correspondence that has been extensively explored in both theoretical~\cite{Coullet1989, Pomeau1993b, Leboeuf2010, Carusotto2014, Larre2015a, Larre2015b, Larre2016, Larre2018, BardonBrun2020, Rodrigues2020} and experimental~\cite{Wan2007, Vocke2016, Michel2018, Vocke2018, Situ2020, Fontaine2020, Eloy2021, Abuzarli2022, BakerRasooli2023, Glorieux2025} studies. An advantage of such \textit{binary superfluids of light} is their accessibility at room temperature in tabletop setups, in contrast to the demanding ultracold engineering needed for atomic superfluids.

Motivated by seminal theoretical studies~\cite{Martone2021, Martone2023}, a recent experimental work~\cite{Piekarski2025} has reported the observation of the density and spin modes in a binary superfluid of light within the miscible repulsive regime, where intracomponent repulsion dominates over intercomponent repulsion~\cite{Pitaevskii2016}. The authors have measured the dispersion relations of these collective excitations, which take the form of density and spin Bogoliubov phonons~\cite{Pitaevskii2016}. Notably, in this system, the corresponding speeds of sound have been found to display an unexpected inversion due to saturation of the optical nonlinearity. As a result, Landau's criterion for \textit{superfluidity}---dissipationless flow---may apply to the density mode rather than the spin mode, contrary to predictions based on the two-component Gross-Pitaevskii equation in the miscible repulsive phase~\cite{Pitaevskii2016, Kim2020, Kim2021, Cominotti2022}.

Based on previous theoretical work~\cite{Pavloff2002, Astrakharchik2004, Frisch1992, Pomeau1993a, Josserand1997, Josserand1999, Rica2001, Pinsker2014, Pigeon2021, Huynh2024a, Huynh2024b}, the present study thoroughly investigates the \textit{critical speed} below which this intrinsically two-dimensional (2D) system exhibits superfluid flow past a localized obstacle. It also provides a qualitative overview of the obstacle-induced fluid's excitations just above this speed. Crucially, the obstacle is taken to be polarization-sensitive, enabling it to excite not only the density mode but also the spin mode of the superfluid, and its parameters are explored across a wide range. While our analysis fully accounts for the optical saturation effects observed in Ref.~\cite{Piekarski2025}, in the weak-saturation regime, it also formally applies to Bose-Bose superfluid mixtures realized with ultracold atoms.

The paper is structured as follows. Section~\ref{Sec:Model} introduces the hydrodynamic equations governing the system, focusing on the total and relative densities and velocities of the binary superfluid of light. Section~\ref{Sec:Lin} then analyzes the system's flow in the regime where the obstacle potential acts as a weak perturbation compared to the interaction energy. This approach allows for the determination of the critical speed based on Landau's criterion, applied to the density and spin Bogoliubov modes identified in Ref.~\cite{Piekarski2025}. The corresponding onset of dissipation, resulting from excitation of these modes, is investigated through the drag force exerted by the superfluid on the obstacle. Section~\ref{Sec:Nonlin} goes beyond perturbation theory to characterize the critical speed as a function of the obstacle's amplitudes. This is done for an obstacle of large spatial extent, by analyzing the conditions for strong ellipticity of the stationary flow equations within the hydraulic and incompressible approximations. This section also explores the breakdown of superfluidity in this regime through real-time numerical simulations, revealing that dissipation is triggered by the emission of vortex and soliton structures in both components. Finally, Sec.~\ref{Sec:Outro} summarizes the main conclusions and outlines prospects for future research.

\section{Model equations}
\label{Sec:Model}

In this section, we first describe the optical setup of interest, followed by the introduction of the wave equation that governs the dynamics of the binary superfluid of light. We then derive a hydrodynamic formulation for it and, finally, establish the model describing the flow incident on the obstacle.

\subsection{Binary superfluid}
\label{SubSec:BinarySF}

We consider a polarized, monochromatic laser beam propagating along the \maths{z} axis of a nonlinear medium. In the basis of left- (\maths{+}) and right- (\maths{-}) circular polarizations, \maths{\hat{\boldsymbol{\sigma}}_{\pm}=(\hat{\mathbf{x}}\pm i\hat{\mathbf{y}})/\sqrt{2}}, its complex-valued electric field in the transverse \maths{x{-}y} plane can be expressed as~\cite{Boyd2020}
\begin{equation}
\label{Eq:E}
\mathbf{E}=(\psi_{+}\hat{\boldsymbol{\sigma}}_{+}+\psi_{-}\hat{\boldsymbol{\sigma}}_{-})e^{i(kz-\omega t)},
\end{equation}
where \maths{\omega/(2\pi)} denotes the laser frequency and \maths{k} is the propagation wave number. The two amplitudes \maths{\psi_{+}(\mathbf{r}=x\hat{\mathbf{x}}+y\hat{\mathbf{y}},z)} and \maths{\psi_{-}(\mathbf{r},z)} are generally different, so that~\eqref{Eq:E} accounts for any possible polarization of the laser. The nonlinear medium in which the beam propagates is characterized by the local refractive index
\begin{equation}
\label{Eq:n+-}
n_{\pm}=n_{0}+n_{1,\pm}(\mathbf{r})+\frac{n_{2}I_{\pm}+n_{2,+-}I_{\mp}}{1+I/I_{\mathrm{sat}}}.
\end{equation}
In this equation, \maths{n_{0}} is a uniform background contribution, which defines the wave number \maths{k=n_{0}\omega/c_{0}}, with \maths{c_{0}} the vacuum speed of light. The linear contribution \maths{n_{1,\pm}(\mathbf{r})} accounts for birefringence~\cite{Boyd2020}, i.e., it depends on the polarization state of the field~\eqref{Eq:E}, and is also assumed to depend on the transverse spatial coordinates \maths{x} and \maths{y}. In practice, such a spatial dependence can be achieved using an intense auxiliary laser beam of waist \maths{w}, with both negligible diffraction and (de)focusing as it propagates along the \maths{z} axis~\cite{Michel2018, Eloy2021}. The contribution \maths{n_{1,\pm}(\mathbf{r})} then typically falls rapidly to zero for distances \maths{r=|\mathbf{r}|} larger than \maths{w}. The last term on the right-hand side of Eq.~\eqref{Eq:E} is the nonlinear contribution to the refractive index (also birefringent in general), as characterized in Ref.~\cite{Piekarski2025}. Here, \maths{n_{2}=n_{2,++}=n_{2,--}} and \maths{n_{2,+-}} are the two independent nonlinear-refractive-index coefficients of the medium, \maths{I_{\pm}=n_{0}\epsilon_{0}c_{0}\rho_{\pm}/2} denotes the local intensity of the \maths{\pm} wave, with \maths{\rho_{\pm}=|\psi_{\pm}|^{2}} and \maths{\epsilon_{0}} the vacuum permittivity, and \maths{I=I_{+}+I_{-}} is the total intensity of the laser. The \maths{I/I_{\mathrm{sat}}} dependence of the denominator accounts for the saturation of the optical nonlinearity observed in Ref.~\cite{Piekarski2025} when \maths{I} gets larger than some intensity \maths{I_{\mathrm{sat}}}~\cite{Michel2018, Eloy2021, Glorieux2025}. When \maths{I\ll I_{\mathrm{sat}}}, the nonlinear contribution is reduced to the usual Kerr response \maths{n_{2}I_{\pm}+n_{2,+-}I_{\mp}} of the medium to the optical field~\eqref{Eq:E}~\cite{Boyd2020}.

The laser amplitudes \maths{\psi_{+}(\mathbf{r},z)} and \maths{\psi_{-}(\mathbf{r},z)} in~\eqref{Eq:E} are, by nature, slowly varying functions of \maths{z}. They satisfy, in the paraxial approximation~\cite{Boyd2020} and for the refractive index~\eqref{Eq:n+-}, the inhomogeneous nonlinear Schr\"odinger-type equation
\begin{equation}
\label{Eq:NLS}
i\partial_{z}\psi_{\pm}=\bigg[{-}\frac{\nabla^{2}}{2k}+U_{\pm}(\mathbf{r})+\frac{g\rho_{\pm}+g_{+-}\rho_{\mp}}{1+\rho/\rho_{\mathrm{sat}}}\bigg]\psi_{\pm},
\end{equation}
where \maths{\nabla=\partial_{\mathbf{r}}} and we have defined \maths{U_{\pm}(\mathbf{r})=-kn_{1,\pm}(\mathbf{r})/n_{0}}, \maths{(g,g_{+-})=-k(n_{2},n_{2,+-})\epsilon_{0}c_{0}/2}, \maths{\rho=\rho_{+}+\rho_{-}}, and \maths{\rho_{\mathrm{sat}}=2I_{\mathrm{sat}}/(n_{0}\epsilon_{0}c_{0})}. By analogy with the two-component Gross-Pitaevskii equation that describes the mean-field dynamics of a Bose-Bose superfluid mixture~\cite{Pitaevskii2016}, Eq.~\eqref{Eq:NLS} can be interpreted as governing the evolution, in the effective time variable \maths{z}, of the wave function \maths{\psi_{\pm}=\sqrt{\rho_{\pm}}\exp(i\phi_{\pm})} of the \maths{\pm} component of a 2D binary superfluid of photons of mass \maths{k}. Here, \maths{\rho_{\pm}} and \maths{\mathbf{v}_{\pm}=\nabla\phi_{\pm}/k} represent the local and instantaneous density and velocity, respectively, of each component in the \maths{x{-}y} plane. The potential \maths{U_{\pm}(\mathbf{r})} describes a localized, static obstacle affecting the flow of the \maths{\pm} component. This potential is centered at \maths{r=0} and vanishes for distances \maths{r\gg w}. For concreteness, we model it as a disk of radius \maths{w} and constant amplitude \maths{U_{\pm}}:
\begin{equation}
\label{Eq:U+-}
U_{\pm}(\mathbf{r})=U_{\pm}\Theta(w-r),
\end{equation}
where \maths{\Theta} is the Heaviside step function. A comparison of Refs.~\cite{Huynh2024a, Huynh2024b} with~\cite{Kwon2015, Kwak2023} shows that this simple modeling is well suited to describe realistic situations [where the obstacle potential \maths{U_{\pm}(\mathbf{r})\simeq U_{\pm}\exp(-r^{2}/w^{2})} is, e.g., closer to a Gaussian function] while providing a comprehensive analytical framework for the critical speed. In this article, we restrict our study to a repulsive obstacle potential, corresponding to \maths{U_{\pm}>0}. The sum of its two components \maths{U_{+}(\mathbf{r})} and \maths{U_{-}(\mathbf{r})} and the difference between the two are, respectively,
\begin{equation}
\label{Eq:U}
U(\mathbf{r})=U\Theta(w-r),\quad U=U_{+}+U_{-}>0,
\end{equation}
and
\begin{equation}
\label{Eq:u}
u(\mathbf{r})=u\Theta(w-r),\quad u=U_{+}-U_{-}\in(-U,U).
\end{equation}
Finally, the nonlinear term on the right-hand side of Eq.~\eqref{Eq:NLS} describes intra- (\maths{g}) and intercomponent (\maths{g_{+-}}) interactions. In this work, we assume that \maths{g>g_{+-}>0}, which is the miscible repulsive regime experimentally investigated in Ref.~\cite{Piekarski2025}.

To ensure concise notations, from now on we work with dimensionless units, defined with respect to a reference density \maths{\rho_{\mathrm{ref}}} that will be specified in Sec.~\ref{SubSec:Hydro}. Specifically, we normalize densities by \maths{\rho_{\mathrm{ref}}}, energies by \maths{\mu=g\rho_{\mathrm{ref}}}, time variables by \maths{1/\mu}, position variables by \maths{\xi=1/(k\mu)^{1/2}}, and velocities by \maths{c=1/(k\xi)}. In these units, Eq.~\eqref{Eq:NLS} becomes
\begin{equation}
\label{Eq:dNLS}
i\partial_{z}\psi_{\pm}=\bigg[{-}\frac{\nabla^{2}}{2}+U_{\pm}(\mathbf{r})+\frac{\rho_{\pm}+\alpha\rho_{\mp}}{1+\beta\rho}\bigg]\psi_{\pm},
\end{equation}
where the coefficient
\begin{equation}
\label{Eq:alpha}
\alpha=\frac{g_{+-}}{g}\in(0,1)
\end{equation}
quantifies intercomponent interactions relative to intracomponent interactions, and
\begin{equation}
\label{Eq:beta}
\beta=\frac{\rho_{\mathrm{ref}}}{\rho_{\mathrm{sat}}}
\end{equation}
determines the saturation degree of the optical nonlinearity. Assuming \maths{\rho_{\mathrm{ref}}\neq\rho_{\mathrm{sat}}}, the limits \maths{\beta\ll1} and \maths{\beta\gg1} correspond to the Kerr and saturated regimes, respectively.

\subsection{Hydrodynamic description}
\label{SubSec:Hydro}

The wave equation~\eqref{Eq:dNLS} admits a convenient hydrodynamic representation, based on the following parametrization of \maths{\psi_{+}} and \maths{\psi_{-}} (see, e.g., Refs.~\cite{Kasamatsu2005, Kamchatnov2014}):
\begin{equation}
\label{Eq:psi+-}
\begin{bmatrix}
\psi_{+} \\ \psi_{-}
\end{bmatrix}
=\sqrt{\rho}e^{i\Phi/2}
\begin{bmatrix}
\cos(\theta/2)e^{i\phi/2}e^{-i\mu_{+}z} \\ \sin(\theta/2)e^{-i\phi/2}e^{-i\mu_{-}z}
\end{bmatrix}
.
\end{equation}
This representation is equivalent to parametrizing the \maths{+} and \maths{-} densities as \maths{\rho_{+}=\rho\cos^{2}(\theta/2)} and \maths{\rho_{-}=\rho\sin^{2}(\theta/2)}, and the phases as \maths{\phi_{\pm}=(\Phi\pm\phi)/2-\mu_{\pm}z}. From this, we obtain the following combinations:
\begin{equation}
\label{Eq:TotRel}
\begin{dcases}
\rho_{+}+\rho_{-}=\rho \\
\mathbf{v}_{+}+\mathbf{v}_{-}=\nabla\Phi\equiv2\mathbf{V}
\end{dcases}
,~~~
\begin{dcases}
\rho_{+}-\rho_{-}=\rho\cos\theta\equiv\sigma \\
\mathbf{v}_{+}-\mathbf{v}_{-}=\nabla\phi\equiv\mathbf{v}
\end{dcases}
.
\end{equation}
These relations allow for a clear physical interpretation of the various fields: \maths{\rho(\mathbf{r},z)} is the total density of the binary superfluid (as already introduced in Sec.~\ref{SubSec:BinarySF}); \maths{\mathbf{V}(\mathbf{r},z)} and \maths{\Phi(\mathbf{r},z)} correspond to the mean (or center-of-mass) velocity and the total-velocity potential; \maths{\sigma(\mathbf{r},z)} and \maths{\theta(\mathbf{r},z)} describe the relative (or spin) density (such that \maths{|\sigma|/\rho<1}) and the ``magnetization'' angle; and \maths{\mathbf{v}(\mathbf{r},z)} and \maths{\phi(\mathbf{r},z)} are the relative velocity and the corresponding potential. Finally, \maths{\mu_{\pm}} denotes the chemical potential of the \maths{\pm} component. Insertion of definitions~\eqref{Eq:psi+-} and~\eqref{Eq:TotRel} into Eq.~\eqref{Eq:dNLS} results in the following, coupled hydrodynamiclike equations for the real fields \maths{\rho}, \maths{\Phi}, \maths{\theta}, and \maths{\phi}:
\begin{align}
\label{Eq:rho}
\partial_{z}\rho&\left.=-\nabla\cdot(\rho\mathbf{V})-\frac{\nabla\cdot(\rho\mathbf{v}\cos\theta)}{2},\right. \\
\notag
\partial_{z}\Phi&\left.=\mu_{+}+\mu_{-}-\frac{(1+\alpha)\rho}{1+\beta\rho}-U(\mathbf{r})-V^{2}-\frac{v^{2}}{4}\right. \\
\label{Eq:Phi}
&\left.\hphantom{=}+\frac{\nabla^{2}\sqrt{\rho}}{\sqrt{\rho}}+\frac{\nabla\cdot(\rho\nabla\theta)}{2\rho\tan\theta}-\frac{(\nabla\theta)^{2}}{4},\right. \\
\label{Eq:theta}
\partial_{z}\theta&\left.=-\mathbf{V}\cdot\nabla\theta+\frac{\nabla\cdot(\rho\mathbf{v}\sin\theta)}{2\rho},\right. \\
\notag
\partial_{z}\phi&\left.=\mu_{+}-\mu_{-}-\frac{(1-\alpha)\rho\cos\theta}{1+\beta\rho}-u(\mathbf{r})-\mathbf{V}\cdot\mathbf{v}\right. \\
\label{Eq:phi}
&\left.\hphantom{=}-\frac{\nabla\cdot(\rho\nabla\theta)}{2\rho\sin\theta}.\right.
\end{align}
It is worth noting that Eqs.~\eqref{Eq:rho} and~\eqref{Eq:theta} can be reformulated in the compact form
\begin{equation}
\label{Eq:Cont}
\partial_{z}\!
\begin{bmatrix}
\rho \\ \sigma
\end{bmatrix}
=-\nabla\cdot\bigg(\!
\begin{bmatrix}
\rho \\ \sigma
\end{bmatrix}
\mathbf{V}+
\begin{bmatrix}
\sigma \\ \rho
\end{bmatrix}
\frac{\mathbf{v}}{2}\bigg),
\end{equation}
which will be used in Sec.~\ref{Sec:Nonlin}. Equation~\eqref{Eq:Cont} gives the continuity equations for the total and relative densities, while Eqs.~\eqref{Eq:Phi} and~\eqref{Eq:phi} are the unsteady Bernoulli equations for the corresponding velocity potentials.

\begin{figure}[t!]
\begin{center}
\includegraphics[width=\linewidth]{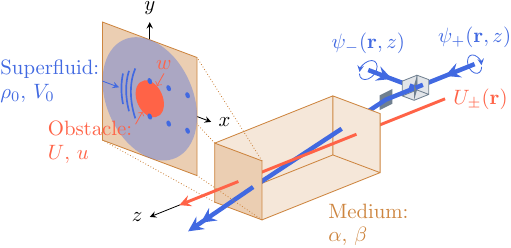}
\end{center}
\caption{By slightly tilting uniform, equally bright left- (\maths{+}) and right- (\maths{-}) circularly polarized beams with same angle with respect to the \maths{z} axis (blue arrows), we create, in the transverse \maths{x{-}y} plane, a fully balanced binary superflow of light with total density \maths{\rho_{0}} and mean velocity \maths{V_{0}}, here along the positive-\maths{x} direction (blue spot in the zoom). This 2D flow impinges on a birefringent optical obstacle of radius \maths{w}, characterized by total and relative potential amplitudes \maths{U} and \maths{u} (red spot in the zoom). The obstacle can be created using an auxiliary beam along the \maths{z} axis (red arrow). The medium (brown box) through which the lasers propagate provides \maths{\pm}/\maths{\pm} and \maths{+}/\maths{-} interactions (\maths{\alpha}) which saturate (\maths{\beta}) at high laser intensity.}
\label{Fig:Setup}
\end{figure}

In what follows, we consider a flow that is homogeneous (\maths{\mathbf{r}} independent) and stationary (\maths{z} independent) at large distance from the obstacle in the \maths{x{-}y} plane. We denote the constant total density of this flow by \maths{\rho_{0}}, and choose it, in proper units, as the reference density introduced in Sec.~\ref{SubSec:BinarySF}: \maths{\rho_{\mathrm{ref}}=\rho_{0}}. Thus, in dimensionless units, we have \maths{\rho_{0}=1}. Furthermore, its constant mean velocity is taken to be \maths{\mathbf{V}_{0}=V_{0}\hat{\mathbf{x}}}, pointing along the positive-\maths{x} direction without loss of generality. Finally, we consider the simplest case where both its relative density and velocity vanish, i.e., \maths{\sigma_{0}=0} (corresponding to equal component densities: \maths{\rho_{\pm,0}=1/2}) and \maths{\mathbf{v}_{0}=0} (equal component velocities: \maths{\mathbf{v}_{\pm,0}=\mathbf{V}_{0}}), respectively. Under these conditions, far from the obstacle, the fields approach the asymptotic configuration
\begin{equation}
\label{Eq:AC}
\begin{bmatrix}
\rho \\ \Phi \\ \theta \\ \phi
\end{bmatrix}
\underset{r\gg\max\{1,w\}}{\simeq}
\begin{bmatrix}
1 \\ 2V_{0}x \\ \pi/2 \\ 0
\end{bmatrix}
,
\end{equation}
up to irrelevant additive constants for the potentials \maths{\Phi} and \maths{\phi}, and an inconsequential congruence modulo \maths{\pi} for the angle \maths{\theta}. Solution~\eqref{Eq:AC} to Eqs.~\eqref{Eq:rho}--\eqref{Eq:phi} fixes the value of the chemical potentials: \maths{\mu_{\pm}=(\mu_{0}+V_{0}{}^{2})/2}, where
\begin{equation}
\label{Eq:mu0}
\mu_{0}=\frac{1+\alpha}{1+\beta}.
\end{equation}
As demonstrated numerically~\cite{Carusotto2014, Larre2015a} and experimentally~\cite{Michel2018, Eloy2021} in the case of a single-component superfluid of light, such a 2D stationary state is expected to be spontaneously reached after some propagation time \maths{z} from the entrance of the medium. In practice, in a manner similar to Refs.~\cite{Michel2018, Eloy2021}, it can be realized by superimposing two wide, top-hat-shaped \maths{+} and \maths{-} polarized beams with  equal incident intensities, \maths{I_{\pm,0}\equiv n_{0}\epsilon_{0}c_{0}\rho_{0}/4}, and  tilting them slightly by the same incidence angle with respect to the \maths{z} axis (here along the positive-\maths{x} direction), \maths{\alpha_{\pm,0}\equiv\arctan(|\nabla\phi_{\pm,0}|/k)=\arctan V_{0}\simeq V_{0}\ll1}.\footnote{Despite this finite \maths{\alpha_{\pm,0}}, we keep the carrier propagation along the \maths{z} axis and the envelope polarization in the \maths{x{-}y} plane. This is legitimate in the limit \maths{\alpha_{\pm,0}\simeq V_{0}\ll1} that underlies Eq.~\eqref{Eq:NLS}. Corrections to this assumption are of higher order and lie beyond the paraxial description of the superfluid of light~\cite{Martone2021, Martone2023}.} This configuration, including the auxiliary beam creating the obstacle potential \maths{U_{\pm}(\mathbf{r})}, is illustrated in Fig.~\ref{Fig:Setup}.

In Sec.~\ref{Sec:Lin}, we study the condition for superfluidity of this flow when the influence of the obstacle on the fluid can be treated perturbatively. In Sec.~\ref{Sec:Nonlin}, we extend the analysis beyond this perturbative regime.

\section{Linear response}
\label{Sec:Lin}

As a first approach, we analyze how the flow behaves in the presence of the obstacle within linear-response theory, which applies when the amplitude \maths{U_{\pm}} of the potential~\eqref{Eq:U+-} is much smaller than the typical interaction energy \maths{\mu_{0}} defined in Eq.~\eqref{Eq:mu0}.

\subsection{Landau's criterion}
\label{SubSec:Landau}

In this regime, the obstacle induces slight deviations \maths{\delta\rho(\mathbf{r},z)}, \maths{\delta\Phi(\mathbf{r},z)}, \maths{\delta\theta(\mathbf{r},z)}, and \maths{\delta\phi(\mathbf{r},z)=O(U_{\pm}/\mu_{0})} from the reference fields in~\eqref{Eq:AC}, and linearization of the flow equations~\eqref{Eq:rho}--\eqref{Eq:phi} with respect to these fluctuations leads to, in Fourier space,
\begin{equation}
\label{Eq:BdG}
(i\Omega_{0}\mathbb{1}+\mathbb{L})
\begin{bmatrix}
\delta\tilde{\rho} \\ \delta\tilde{\Phi} \\ \delta\tilde{\theta} \\ \delta\tilde{\phi}
\end{bmatrix}
=2\pi
\begin{bmatrix}
0 \\ \tilde{U}_{\mathbf{K}} \\ 0 \\ \tilde{u}_{\mathbf{K}}
\end{bmatrix}
\delta(\Omega-i0^{+}),
\end{equation}
where \maths{\mathbb{1}} is the identity matrix and
\begin{equation}
\label{Eq:L}
\mathbb{L}=
\begin{bmatrix}
0 & \frac{K^{2}}{2} & 0 & 0 \\
-\frac{K^{2}}{2}-\frac{1+\alpha}{(1+\beta)^{2}} & 0 & 0 & 0 \\
0 & 0 & 0 & -\frac{K^{2}}{2} \\
0 & 0 & \frac{K^{2}}{2}+\frac{1-\alpha}{1+\beta} & 0
\end{bmatrix}
.
\end{equation}
In Eq.~\eqref{Eq:BdG}, \maths{\tilde{f}=\int d^{2}\mathbf{r}dz~f\exp[-i(\mathbf{K}\cdot\mathbf{r}-\Omega z)]} denotes the Fourier transform, \maths{\Omega_{0}=\Omega-V_{0}K_{x}} is the energy of the fluctuations \maths{\delta\tilde{\rho}}, \maths{\delta\tilde{\Phi}}, \maths{\delta\tilde{\theta}}, and \maths{\delta\tilde{\phi}} in the reference frame moving with velocity \maths{\mathbf{V}_{0}}, and the infinitesimal imaginary shift \maths{-i0^{+}} is introduced so that the response of the fluid to the presence of the obstacle is causal. This amounts to slowly branching the obstacle potential as \maths{U_{\pm}(\mathbf{r})\equiv U_{\pm}(\mathbf{r},z)=U_{\pm}(\mathbf{r})\exp(\eta z)} with \maths{\eta\to0^{+}}~\cite{Pines1966}.

The block-diagonal nature of the matrix \maths{\mathbb{L}} defined in Eq.~\eqref{Eq:L} indicates that the linearized fluctuations of the fluid correspond to two independent excitation modes. The first mode, which is of density type and hereafter indicated by the subscript ``\maths{\mathrm{d}},'' corresponds to fluctuations in the total fields \maths{\rho} and \maths{\Phi}, and the second one, which is of spin (``\maths{\mathrm{s}}'') type, corresponds to fluctuations in the relative fields \maths{\theta} and \maths{\phi}~\cite{Pitaevskii2016}. Their respective dispersion relations, obtained from \maths{\det(i\Omega_{0}\mathbb{1}+\mathbb{L})=0}, are given by \maths{\Omega_{0}=\pm\Omega_{\mathrm{d}}} and \maths{\Omega_{0}=\pm\Omega_{\mathrm{s}}}, where
\begin{equation}
\label{Eq:Omegads}
\Omega_{\mathrm{d},\mathrm{s}}=\sqrt{\frac{K^{2}}{2}\bigg(\frac{K^{2}}{2}+2c_{\mathrm{d},\mathrm{s}}{}^{2}\bigg)}.
\end{equation}
They are of Bogoliubov type~\cite{Pitaevskii2016}, quadratic (particlelike) at large \maths{K} and linear (phononlike) at small \maths{K}, with respective absolute speeds of sound
\begin{equation}
\label{Eq:cdcs}
c_{\mathrm{d}}=\sqrt{\frac{1+\alpha}{2}}\frac{1}{1+\beta},\quad c_{\mathrm{s}}=\sqrt{\frac{1-\alpha}{2}}\frac{1}{\sqrt{1+\beta}}
\end{equation}
in the reference frame moving with velocity \maths{\mathbf{V}_{0}}.

The Bogoliubov speed of sound \maths{c_{\mathrm{d}(\mathrm{s})}} in Eq.~\eqref{Eq:Omegads} coincides with Landau's critical speed for energetic stability of the fluid with respect to the density (spin) mode~\cite{Pitaevskii2016}:
\begin{equation}
\label{Eq:Landau}
V_{0}<\min_{\mathbf{K}}\frac{\Omega_{\mathrm{d}(\mathrm{s})}}{K}=c_{\mathrm{d}(\mathrm{s})}.
\end{equation}
When this condition is met, the density (spin) mode is never stimulated, and the total (relative) density and velocity of the fluid remain nearly unperturbed by the obstacle. In contrast, when condition~\eqref{Eq:Landau} is not satisfied, the fluid dissipates energy through the emission of density (spin) Bogoliubov waves on top of background~\eqref{Eq:AC}. The Landau condition for full energetic stability---and thus superfluidity---is therefore, within the present linear-response approach,
\begin{equation}
\label{Eq:VcLandau}
V_{0}<V_{\mathrm{c}}=\min\{c_{\mathrm{d}},c_{\mathrm{s}}\}.
\end{equation}
For the sake of completeness, in Appendix~\ref{App:Imbalanced} we also give the Landau critical speed \maths{V_{\mathrm{c}}} in the two more general cases where \maths{\sigma_{0}=0} and \maths{\mathbf{v}_{0}\neq0} (which corresponds to a superfluid that is imbalanced in velocity far from the obstacle: \maths{\mathbf{v}_{+,0}\neq\mathbf{v}_{-,0}}), and where \maths{\sigma_{0}\neq0} and \maths{\mathbf{v}_{0}=0} (superfluid imbalanced in density: \maths{\rho_{+,0}\neq\rho_{-,0}}).

\begin{figure}[t!]
\begin{center}
\includegraphics[width=\linewidth]{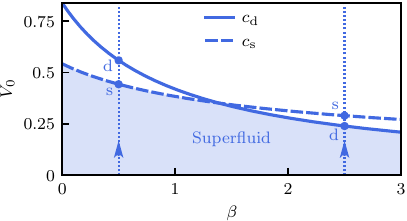}
\end{center}
\caption{Density speed of sound \maths{c_{\mathrm{d}}} [first equation in~\eqref{Eq:cdcs}; solid curve] and spin speed of sound \maths{c_{\mathrm{s}}} (second equation; dashed curve) as functions of the saturation parameter \maths{\beta} for the intercomponent interaction constant \maths{\alpha\simeq0.41} of Ref.~\cite{Piekarski2025}. The coordinates of the intersection point are \maths{\beta=2\alpha/(1-\alpha)\simeq1.39} and \maths{c_{\mathrm{d},\mathrm{s}}=(1-\alpha)/[2(1+\alpha)]^{1/2}\simeq0.35}. At low \maths{\beta}, in the Kerr regime, \maths{c_{\mathrm{d}}\simeq[(1+\alpha)/2]^{1/2}\simeq0.84} and \maths{c_{\mathrm{s}}\simeq[(1-\alpha)/2]^{1/2}\simeq0.54}. The shaded area represents the parameter regime where condition~\eqref{Eq:VcLandau} holds and the two-component flow of mean asymptotic velocity \maths{V_{0}} is superfluid, within perturbation theory. The arrows indicate the opening of the density (``\maths{\mathrm{d}}'') and spin (``\maths{\mathrm{s}}'') dissipation channels as \maths{V_{0}} is increased from the superfluid phase at fixed \maths{\beta}'s below (\maths{\beta=0.5}; \maths{V_{\mathrm{c}}=c_{\mathrm{s}}\simeq0.44<c_{\mathrm{d}}\simeq0.56}) and above (\maths{\beta=2.5}; \maths{V_{\mathrm{c}}=c_{\mathrm{d}}\simeq0.24<c_{\mathrm{s}}\simeq0.29}) the curve inversion at \maths{\beta\simeq1.39}.}
\label{Fig:Landau}
\end{figure}

We show in Fig.~\ref{Fig:Landau} the behavior of the two speeds of sound \maths{c_{\mathrm{d}}} and \maths{c_{\mathrm{s}}} as a function of the saturation parameter \maths{\beta=1/\rho_{\mathrm{sat}}}. In the Kerr regime \maths{\beta\ll1}, the \maths{\beta} contributions in the denominators of Eqs.~\eqref{Eq:cdcs} can be neglected, yielding the simplified expressions \maths{c_{\mathrm{d}}\simeq[(1+\alpha)/2]^{1/2}} and \maths{c_{\mathrm{s}}\simeq[(1-\alpha)/2]^{1/2}}. These are the well-known Bogoliubov speeds of sound  for the density and spin modes, respectively, in the framework of the two-component Gross-Pitaevskii equation (without Rabi coupling between \maths{\psi_{+}} and \maths{\psi_{-}})~\cite{Pitaevskii2016}. In this regime, \maths{c_{\mathrm{s}}<c_{\mathrm{d}}}, so condition~\eqref{Eq:VcLandau} is limited by the spin mode, with a critical speed \maths{V_{\mathrm{c}}=c_{\mathrm{s}}}. As the saturation parameter \maths{\beta} increases, \maths{c_{\mathrm{d}}} decreases faster than \maths{c_{\mathrm{s}}}, and for \maths{\beta>2\alpha/(1-\alpha)}, \maths{c_{\mathrm{d}}} becomes smaller than \maths{c_{\mathrm{s}}}. This inversion of the sound speeds was first reported in the experiment of Ref.~\cite{Piekarski2025}. When it occurs, the critical speed switches to \maths{V_{\mathrm{c}}=c_{\mathrm{d}}}. In Fig.~\ref{Fig:Landau}, we highlight as a shaded area the region in which the superfluidity condition~\eqref{Eq:VcLandau} is fulfilled, using the parameter value \maths{\alpha\simeq0.41} adopted in Ref.~\cite{Piekarski2025}. For this value, the  speed-of-sound inversion occurs at \maths{\beta\simeq1.39}. It is also worth noting that \maths{c_{\mathrm{s}}\to0} in the limit \maths{\alpha\to1^{-}}, often referred to as the SU(2) or Manakov limit~\cite{Rabec2025}. In this case, the spin branch \maths{\Omega_{\mathrm{s}}\simeq K^{2}/2} of the Bogoliubov dispersion relation becomes purely quadratic rather than linear at low \maths{K}, and the flow is consequently always dissipative according to Landau's criterion. This regime marks the onset of large relative fluctuations leading to phase separation~\cite{Pitaevskii2016}, and thus lies beyond the scope of the present work.

\subsection{Drag force}
\label{SubSec:Drag}

A natural way to characterize the superfluidity of the flow and the opening of the density and spin dissipation channels at \maths{c_{\mathrm{d}}} and \maths{c_{\mathrm{s}}} is to compute the drag force \maths{\mathbf{F}} exerted by the fluid on the obstacle: a nonzero \maths{\mathbf{F}} implies dissipation through the emission of density, spin, or both density and spin excitations, whereas \maths{\mathbf{F}=0} corresponds to a superfluid flow. We define this drag force in a manner similar to that in, e.g., Refs.~\cite{Pavloff2002, Astrakharchik2004}: \maths{\mathbf{F}=\langle\nabla\mathbb{U}(\mathbf{r})\rangle}, where \maths{\mathbb{U}(\mathbf{r})=\mathrm{diag}[U_{+}(\mathbf{r}),U_{-}(\mathbf{r})]} represents the obstacle potential and \maths{\langle f\rangle=\int d^{2}\mathbf{r}~\langle\psi|f|\psi\rangle} denotes the average over the superfluid wave function \maths{|\psi\rangle=\textsuperscript{t}[\psi_{+}~\psi_{-}]}. This gives, integrating by parts,
\begin{equation}
\label{Eq:F}
\mathbf{F}=-\int d^{2}\mathbf{r}~[U_{+}(\mathbf{r})\nabla\rho_{+}+U_{-}(\mathbf{r})\nabla\rho_{-}].
\end{equation}
Interestingly, one can show~\cite{Larre2015a} that~\eqref{Eq:F} is proportional to the genuine radiation force~\cite{Barnett2006, Loudon2006} experienced by the medium (at low optical nonlinearity).

Expanding \maths{\rho_{\pm}=(1+\delta\rho\mp\delta\theta)/2} to first order in the density fluctuations \maths{\delta\rho} and \maths{\delta\theta}, we get \maths{\mathbf{F}=-\int d^{2}\mathbf{r}~[U(\mathbf{r})\nabla\delta\rho-u(\mathbf{r})\nabla\delta\theta]/2}, in which we can insert the respective expressions of \maths{\delta\rho} and \maths{\delta\theta} deduced from Eq.~\eqref{Eq:BdG}:\footnote{\label{Foot:PhaseFluct}The phase fluctuations \maths{\delta\Phi} and \maths{\delta\phi} are obtained by substituting \maths{-iV_{0}K_{x}} into the numerator of the Fourier-space response functions in Eqs.~\eqref{Eq:deltarho} and~\eqref{Eq:deltatheta}, respectively.}
\begin{align}
\label{Eq:deltarho}
\delta\rho&=\int\frac{d^{2}\mathbf{K}}{(2\pi)^{2}}~\frac{-K^{2}/2}{\Omega_{\mathrm{d}}{}^{2}-(V_{0}K_{x}-i0^{+})^{2}}\tilde{U}_{\mathbf{K}}e^{i\mathbf{K}\cdot\mathbf{r}}, \\
\label{Eq:deltatheta}
\delta\theta&=\int\frac{d^{2}\mathbf{K}}{(2\pi)^{2}}~\frac{K^{2}/2}{\Omega_{\mathrm{s}}{}^{2}-(V_{0}K_{x}-i0^{+})^{2}}\tilde{u}_{\mathbf{K}}e^{i\mathbf{K}\cdot\mathbf{r}}.
\end{align}
This yields
\begin{align}
\notag
\mathbf{F}&\left.=\frac{i}{4}\int\frac{d^{2}\mathbf{K}}{(2\pi)^{2}}~K^{2}\mathbf{K}\bigg[\frac{|\tilde{U}_{\mathbf{K}}|^{2}}{\Omega_{\mathrm{d}}{}^{2}-(V_{0}K_{x}-i0^{+})^{2}}\right. \\
\label{Eq:FPert}
&\left.\hphantom{=}+\frac{|\tilde{u}_{\mathbf{K}}|^{2}}{\Omega_{\mathrm{s}}{}^{2}-(V_{0}K_{x}-i0^{+})^{2}}\bigg].\right.
\end{align}
Due to parity symmetry of the obstacle potential along the \maths{y} axis, \maths{F_{y}} in Eq.~\eqref{Eq:FPert} is by construction identically zero, so that the drag force \maths{\mathbf{F}=F\hat{\mathbf{x}}} is aligned with the mean asymptotic velocity. We evaluate \maths{F} thanks to the residue theorem and eventually find
\begin{equation}
\label{Eq:FPert_bis}
F=\pi w^{2}[U^{2}I_{\mathrm{d}}\Theta(V_{0}-c_{\mathrm{d}})+u^{2}I_{\mathrm{s}}\Theta(V_{0}-c_{\mathrm{s}})],
\end{equation}
where the integral
\begin{equation}
\label{Eq:Ids}
I_{\mathrm{d},\mathrm{s}}=\int_{c_{\mathrm{d},\mathrm{s}}/V_{0}}^{1}dX~\frac{XJ_{1}(2\sqrt{V_{0}{}^{2}X^{2}-c_{\mathrm{d},\mathrm{s}}{}^{2}}w)^{2}}{\sqrt{(1-X^{2})(V_{0}{}^{2}X^{2}-c_{\mathrm{d},\mathrm{s}}{}^{2})}}
\end{equation}
involves the Fourier transform \maths{2\pi wJ_{1}(Kw)/K} of the Heaviside step function in Eqs.~\eqref{Eq:U} and~\eqref{Eq:u}, \maths{J_{1}} being the Bessel function of the first kind. Similar perturbative calculations of the drag in binary quantum fluids are presented in Refs.~\cite{Fil2005, Flayac2013, Larre2013}.

\begin{figure}[t!]
\begin{center}
\includegraphics[width=\linewidth]{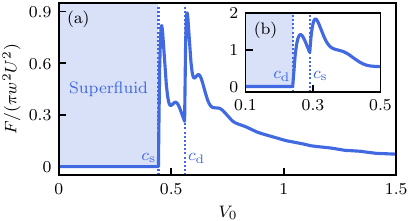}
\end{center}
\caption{Normalized drag force \maths{F/(\pi w^{2}U^{2})} [Eqs.~\eqref{Eq:FPert_bis} and~\eqref{Eq:Ids}] experienced by the obstacle of radius \maths{w} and total and relative potential amplitudes \maths{U} and \maths{u} as a function of the mean asymptotic velocity \maths{V_{0}} of the flow. The two graphs are plotted for the intercomponent interaction constant \maths{\alpha\simeq0.41} of Ref.~\cite{Piekarski2025}, \maths{w=10}, and \maths{|u|=U}. In panel~(a), \maths{\beta=0.5} (left arrow in Fig.~\ref{Fig:Landau}) and \maths{c_{\mathrm{s}}\simeq0.44<c_{\mathrm{d}}\simeq0.56}; in panel~(b), \maths{\beta=2.5} (right arrow) and \maths{c_{\mathrm{d}}\simeq0.24<c_{\mathrm{s}}\simeq0.29}. The shaded region corresponds to the dragless, superfluid regime, which breaks down as soon as \maths{V_{0}} exceeds the critical speed (a)~\maths{V_{\mathrm{c}}=c_{\mathrm{s}}} or (b)~\maths{V_{\mathrm{c}}=c_{\mathrm{d}}}.}
\label{Fig:Drag}
\end{figure}

The drag force~\eqref{Eq:FPert_bis} normalized to \maths{\pi w^{2}U^{2}} is plotted as a function of \maths{V_{0}} in Fig.~\ref{Fig:Drag}. First, one notes that \maths{F=0} when \maths{V_{0}<\min\{c_{\mathrm{d}},c_{\mathrm{s}}\}=V_{\mathrm{c}}}. This is the undeniable signature of superfluid motion, in agreement with Landau's prediction~\eqref{Eq:VcLandau}. Then, \maths{F} undergoes a sudden increase when \maths{V_{0}} crosses \maths{V_{\mathrm{c}}}, and again right above \maths{\max\{c_{\mathrm{d}},c_{\mathrm{s}}\}}. This corresponds to the successive openings of two channels of energy dissipation for the fluid. When \maths{c_{\mathrm{s}}<c_{\mathrm{d}}} as in Fig.~\ref{Fig:Drag}(a) [\maths{c_{\mathrm{d}}<c_{\mathrm{s}}} as in Fig.~\ref{Fig:Drag}(b)], the work \maths{-Fx} imparted to the fluid is dissipated by generating spin (density) waves when \maths{c_{\mathrm{s}}<V_{0}<c_{\mathrm{d}}} (\maths{c_{\mathrm{d}}<V_{0}<c_{\mathrm{s}}}), and then additionally by density (spin) waves when \maths{V_{0}>c_{\mathrm{d}}} (\maths{V_{0}>c_{\mathrm{s}}}). Finally, at larger velocities \maths{V_{0}}, the obstacle potential becomes a weak perturbation compared to the kinetic energy of the fluid. As a result, in the limit \maths{V_{0}\to\infty}, the flow is less and less perturbed by the obstacle and \maths{F\to0}, in agreement with physical intuition.

Assuming the peaks observed in the drag force are well separated from one another, the amplitudes of the \maths{V_{0}\gtrsim c_{\mathrm{d}}} and \maths{V_{0}\gtrsim c_{\mathrm{s}}} peaks are proportional to \maths{U^{2}} and \maths{u^{2}}, respectively [see Eq.~\eqref{Eq:FPert_bis}]. When \maths{u=0}, i.e., \maths{U_{+}=U_{-}}, only the \maths{V_{0}\gtrsim c_{\mathrm{d}}} peak remains, which signals that dissipation occurs solely through the emission of density waves [the spin fluctuations \maths{\delta\theta} and \maths{\delta\phi} are indeed zero in this case; see Eq.~\eqref{Eq:deltatheta} and footnote~\ref{Foot:PhaseFluct}]. This underscores the necessity of a birefringent obstacle potential to excite the spin mode, at least when this obstacle is small.

We also note that the larger the radius \maths{w} of the obstacle, the better resolved these peaks are. Their typical width is indeed proportional to \maths{1/w} [see Eq.~\eqref{Eq:Ids}], so that the peaks are well resolved when \maths{w|c_{\mathrm{d}}-c_{\mathrm{s}}|\gg1}. The observed small oscillations stem from the \maths{J_{1}} function in Eq.~\eqref{Eq:Ids}. These oscillations are absent when a smoothed version of the sharp disk~\eqref{Eq:U+-} is employed, and a fortiori with a standard Gaussian potential. In the limit \maths{w\to0} with \maths{w^{2}U_{\pm}=O(1)}, the obstacle potential~\eqref{Eq:U+-} becomes pointlike and the drag force \maths{F\propto(V_{0}{}^{2}-c_{\mathrm{d},\mathrm{s}}{}^{2})/V_{0}}~\cite{Astrakharchik2004} above \maths{c_{\mathrm{d},\mathrm{s}}} grows almost linearly with \maths{V_{0}}, with distinct slopes depending on whether \maths{V_{0}} is smaller or larger than \maths{\max\{c_{\mathrm{d}},c_{\mathrm{s}}\}}.

Concluding Sec.~\ref{Sec:Lin}, we have identified two mean-velocity thresholds for excitation of the two-component superfluid. Within linear-response theory, these are the speeds \maths{c_{\mathrm{d}}} and \maths{c_{\mathrm{s}}} of Bogoliubov phonons in the density and spin modes. Below the lowest of these two speeds, the flow is superfluid with zero drag. Wave resistance, following a terminology established in the context of capillary-gravity waves~\cite{Raphael1996}, is abruptly activated above this critical speed, resulting in significant drag due to emission of density and spin Bogoliubov waves. However, linear-response theory does not predict how the critical speed depends on the characteristic parameters of the obstacle potential, such as its amplitudes for instance. To explore this dependence, we must employ a theoretical approach that extends beyond. This is addressed in the subsequent section.

\section{Nonlinear response}
\label{Sec:Nonlin}

Hydrodynamically, superfluid motion is characterized by a \textit{stationary} flow that \textit{monotonically flattens} far from the obstacle. In the present section, we derive the conditions for the existence of such a flow in the presence of a repulsive obstacle potential with \textit{arbitrary} total and relative amplitudes \maths{U>0} and \maths{-U<u<U} [see Eqs.~\eqref{Eq:U} and~\eqref{Eq:u}]. In doing so, we go beyond the linear-response regime (\maths{U,|u|\ll\mu_{0}}) investigated in the previous section. 
We show that these conditions reduce to requiring that the mean asymptotic velocity \maths{V_{0}} be bounded by a critical value \maths{V_{\mathrm{c}}}, which depends on \maths{U} and \maths{u}, and can  naturally be identified as the critical speed for superfluidity.

To make the problem analytically tractable, we restrict our attention to an obstacle of radius \maths{w\gg1}, i.e., in proper units, larger than the characteristic healing length \maths{\xi=1/(kg\rho_{0})^{1/2}} introduced in Sec.~\ref{Sec:Model}. In this regime, the hydrodynamic fields vary in the \maths{x{-}y} plane on length scales of the order of \maths{w}, so that all the dispersive, gradient-type terms in Eqs.~\eqref{Eq:Phi} and~\eqref{Eq:phi} can be neglected. This dispersionless approach, sometimes referred to as the \textit{hydraulic} approximation~\cite{Hakim1997, Leszczyszyn2009}, may be regarded as a generalization of the Thomas-Fermi approximation~\cite{Pitaevskii2016} to nonzero flow velocity, and is relevant for experiments~\cite{Michel2018, Eloy2021}.

As previously mentioned, we focus on a stationary (\maths{z}-independent) flow. Correspondingly, the two continuity equations~\eqref{Eq:Cont} read
\begin{equation}
\label{Eq:ContStat}
\nabla\cdot\bigg(\!
\begin{bmatrix}
\rho \\ \sigma
\end{bmatrix}
\mathbf{V}+
\begin{bmatrix}
\sigma \\ \rho
\end{bmatrix}
\frac{\mathbf{v}}{2}\bigg)=
\begin{bmatrix}
0 \\ 0
\end{bmatrix}
,
\end{equation}
and the two Bernoulli equations~\eqref{Eq:Phi} and~\eqref{Eq:phi} become, in the hydraulic approximation,
\begin{align}
\label{Eq:rhoHydrau}
\frac{(1+\alpha)\rho}{1+\beta\rho}&=\mu_{0}-U\Theta(w-r)-(V^{2}-V_{0}{}^{2})-\frac{v^{2}}{4}, \\
\label{Eq:sigmaHydrau}
\frac{(1-\alpha)\sigma}{1+\beta\rho}&=-u\Theta(w-r)-\mathbf{V}\cdot\mathbf{v}.
\end{align}
The  first two equations~\eqref{Eq:ContStat} constitute the differential part of the problem, while the  subsequent two equations~\eqref{Eq:rhoHydrau} and~\eqref{Eq:sigmaHydrau} provide algebraic relations between the densities \maths{\rho}, \maths{\sigma} and the velocities \maths{\mathbf{V}}, \maths{\mathbf{v}} of the binary superfluid. Importantly, we restrict ourselves to hydrodynamic solutions \maths{[\rho(\mathbf{r}),\mathbf{V}(\mathbf{r}),\sigma(\mathbf{r}),\mathbf{v}(\mathbf{r})]} that approach the uniform configuration \maths{(1,V_{0}\hat{\mathbf{x}},0,0)} at a large distance \maths{r\gg w} from the obstacle, as specified in Eq.~\eqref{Eq:AC}, but with the additional requirement that this convergence be monotonic (i.e., typically without damped oscillations). Our aim is to determine the condition under which such superfluid-type solutions exist, expressed in terms of the mean asymptotic velocity \maths{V_{0}} for given obstacle (\maths{U}, \maths{u}) and fluid (\maths{\alpha}, \maths{\beta}) parameters.

\subsection{Elliptic flow}
\label{SubSec:Elliptic}

Fundamentally, such superfluid solutions exist provided the two coupled partial differential equations in~\eqref{Eq:ContStat} are \textit{strongly elliptic}. This condition is satisfied when the slopes \maths{dx/dy} of the characteristic curves \maths{x=x(y)} of these equations (i.e., their characteristic velocities) are purely imaginary. This mathematical property ensures that the solutions of~\eqref{Eq:ContStat} monotonically approach a uniform flow in all directions far from the perturbing obstacle, a hallmark of superfluid motion. First proposed by Frisch, Pomeau, and Rica~\cite{Frisch1992} in the framework of the 2D scalar cubic nonlinear Schr\"odinger equation, this criterion has since become a standard tool~\cite{Pomeau1993a, Josserand1997, Josserand1999, Rica2001, Pinsker2014, Pigeon2021, Huynh2024a, Huynh2024b} which applies to any model of superflow in any dimension of space. 

To implement this criterion in practice, a possibility is to follow Chaplygin's method, which uses the hodograph transform~\cite{Chaplygin1902, Landau1987} to nonperturbatively linearize the continuity equation, thereby allowing the classification of~\eqref{Eq:ContStat} as elliptic. Here, however, we adopt a less demanding yet equally predictive approach, inspired by Refs.~\cite{Josserand1997, Josserand1999}. This method analyzes~\eqref{Eq:ContStat} in the immediate vicinity of the points where the energetic instabilities associated to the breakdown of superfluidity arise. As will be shown numerically in Sec.~\ref{SubSec:Num}, these instabilities grow either at the poles of the obstacle (\maths{x=0} and \maths{y=\pm w^{\pm}}) or within the obstacle itself (\maths{r<w}), where the local velocities of the flow are approximately aligned with the \maths{x} axis:
\begin{equation}
\label{Eq:LocalVv}
(V_{x},V_{y})\simeq(sV,0),\quad(v_{x},v_{y})\simeq(s'v,0),
\end{equation}
with \maths{s=\mathrm{sgn}(V_{x})} and \maths{s'=\mathrm{sgn}(v_{x})}. As will be shown in Sec.~\ref{SubSec:Incomp}, this alignment is exact under the incompressible approximation for the velocity fields.

As a result, in these regions of the plane, the continuity equations~\eqref{Eq:ContStat} simplify to
\begin{align}
\notag
&\left.
\begin{bmatrix}
\rho \\ \sigma
\end{bmatrix}
\partial_{x}V_{x}+sV\partial_{x}\!
\begin{bmatrix}
\rho \\ \sigma
\end{bmatrix}
+
\begin{bmatrix}
\rho \\ \sigma
\end{bmatrix}
\partial_{y}V_{y}\right. \\
\label{Eq:ContStat_2}
&\left.\quad+
\begin{bmatrix}
\sigma \\ \rho
\end{bmatrix}
\frac{\partial_{x}v_{x}}{2}+s'\frac{v}{2}\partial_{x}\!
\begin{bmatrix}
\sigma \\ \rho
\end{bmatrix}
+
\begin{bmatrix}
\sigma \\ \rho
\end{bmatrix}
\frac{\partial_{y}v_{y}}{2}=
\begin{bmatrix}
0 \\ 0
\end{bmatrix}
,\right.
\end{align}
and the scalar product \maths{\mathbf{V}\cdot\mathbf{v}=ss'Vv} in the second of the Bernoulli equations~\eqref{Eq:rhoHydrau} and~\eqref{Eq:sigmaHydrau}. These last two equations show that \maths{\rho} and \maths{\sigma} depend on the magnitudes \maths{V} and \maths{v} of the velocity fields. It then follows that the spatial derivatives of \maths{\rho} and \maths{\sigma} in Eq.~\eqref{Eq:ContStat_2} can be rewritten in terms of derivatives with respect to \maths{V} and \maths{v}, according to \maths{\partial_{x}(\rho,\sigma)=\partial_{V}(\rho,\sigma)\partial_{x}(V_{x}/s)+\partial_{v}(\rho,\sigma)\partial_{x}(v_{x}/s')}. This leads to
\begin{align}
\notag
&\left.\bigg(\!
\begin{bmatrix}
\rho \\ \sigma
\end{bmatrix}
+V\partial_{V}\!
\begin{bmatrix}
\rho \\ \sigma
\end{bmatrix}
+ss'\frac{v}{2}\partial_{V}\!
\begin{bmatrix}
\sigma \\ \rho
\end{bmatrix}
\!\bigg)\partial_{x}V_{x}+
\begin{bmatrix}
\rho \\ \sigma
\end{bmatrix}
\partial_{y}V_{y}\right. \\
\notag
&\left.\quad+\bigg(\!
\begin{bmatrix}
\sigma \\ \rho
\end{bmatrix}
+v\partial_{v}\!
\begin{bmatrix}
\sigma \\ \rho
\end{bmatrix}
+ss'2V\partial_{v}\!
\begin{bmatrix}
\rho \\ \sigma
\end{bmatrix}
\!\bigg)\frac{\partial_{x}v_{x}}{2}+
\begin{bmatrix}
\sigma \\ \rho
\end{bmatrix}
\frac{\partial_{y}v_{y}}{2}\right. \\
\label{Eq:ContStat_3}
&\left.\quad=
\begin{bmatrix}
0 \\ 0
\end{bmatrix}
.\right.
\end{align}
Next, we use Eqs.~\eqref{Eq:rhoHydrau} and~\eqref{Eq:sigmaHydrau} to make explicit the velocity derivatives of \maths{\rho} and \maths{\sigma} in Eq.~\eqref{Eq:ContStat_3}. We find
\begin{align}
\label{Eq:drho_Vv}
&
\begin{dcases}
\frac{\partial_{V}\rho}{\rho}=-\frac{V}{\tilde{c}_{\mathrm{d}}{}^{2}} \\
\frac{\partial_{v}\rho}{\rho}=-\frac{1}{4}\frac{v}{\tilde{c}_{\mathrm{d}}{}^{2}}
\end{dcases}
, \\
\label{Eq:dsigma_Vv}
&
\begin{dcases}
\frac{\partial_{V}\sigma}{\rho}=-ss'\frac{1}{2}\frac{v}{\tilde{c}_{\mathrm{s}}{}^{2}}-\frac{\beta\sigma}{1+\beta\rho}\frac{V}{\tilde{c}_{\mathrm{d}}{}^{2}} \\
\frac{\partial_{v}\sigma}{\rho}=-ss'\frac{1}{2}\frac{V}{\tilde{c}_{\mathrm{s}}{}^{2}}-\frac{1}{4}\frac{\beta\sigma}{1+\beta\rho}\frac{v}{\tilde{c}_{\mathrm{d}}{}^{2}}
\end{dcases}
,
\end{align}
where we have introduced the \emph{local} density and spin speeds of sound
\begin{equation}
\label{Eq:Localcdcs}
\tilde{c}_{\mathrm{d}}=\sqrt{\frac{1+\alpha}{2}}\frac{\sqrt{\rho}}{1+\beta\rho},\quad \tilde{c}_{\mathrm{s}}=\sqrt{\frac{1-\alpha}{2}}\sqrt{\frac{\rho}{1+\beta\rho}}.
\end{equation}
When \maths{\rho=\rho_{0}=1}, these expressions obviously coincide with the uniform expressions~\eqref{Eq:cdcs} for the weakly perturbed superfluid. Substituting Eqs.~\eqref{Eq:drho_Vv} and~\eqref{Eq:dsigma_Vv}, together with the definitions \maths{V_{x,y}=\partial_{x,y}\Phi/2} and \maths{v_{x,y}=\partial_{x,y}\phi}, into Eq.~\eqref{Eq:ContStat_3} finally leads to the following matrix equation for the velocity potentials \maths{\Phi} and \maths{\phi}:
\begin{equation}
\label{Eq:ContPhiphi}
\begin{bmatrix}
A\partial_{xx}+\partial_{yy} & D\partial_{xx}+\frac{\sigma}{\rho}\partial_{yy} \\
C\partial_{xx}+\frac{\sigma}{\rho}\partial_{yy} & B\partial_{xx}+\partial_{yy}
\end{bmatrix}
\begin{bmatrix}
\Phi \\ \phi
\end{bmatrix}
=0,
\end{equation}
where \maths{A}, \maths{B}, \maths{C}, and \maths{D} are expressed as functions of the local densities \maths{\rho}, \maths{\sigma} and velocities \maths{V}, \maths{v} as
\begin{align}
\label{Eq:a}
A&=1-\frac{V^{2}}{\tilde{c}_{\mathrm{d}}{}^{2}}-\frac{1}{4}\frac{v^{2}}{\tilde{c}_{\mathrm{s}}{}^{2}}-ss'\frac{1}{2}\frac{\beta\sigma}{1+\beta\rho}\frac{Vv}{\tilde{c}_{\mathrm{d}}{}^{2}}, \\
\label{Eq:b}
B&=1-\frac{V^{2}}{\tilde{c}_{\mathrm{s}}{}^{2}}-\frac{1}{4}\frac{v^{2}}{\tilde{c}_{\mathrm{d}}{}^{2}}-ss'\frac{1}{2}\frac{\beta\sigma}{1+\beta\rho}\frac{Vv}{\tilde{c}_{\mathrm{d}}{}^{2}}, \\
\label{Eq:c}
C&=\frac{\sigma}{\rho}-\frac{\beta\sigma}{1+\beta\rho}\frac{V^{2}}{\tilde{c}_{\mathrm{d}}{}^{2}}-ss'\frac{1}{2}\bigg(\frac{1}{\tilde{c}_{\mathrm{d}}{}^{2}}+\frac{1}{\tilde{c}_{\mathrm{s}}{}^{2}}\bigg)Vv, \\
\label{Eq:d}
D&=\frac{\sigma}{\rho}-\frac{1}{4}\frac{\beta\sigma}{1+\beta\rho}\frac{v^{2}}{\tilde{c}_{\mathrm{d}}{}^{2}}-ss'\frac{1}{2}\bigg(\frac{1}{\tilde{c}_{\mathrm{d}}{}^{2}}+\frac{1}{\tilde{c}_{\mathrm{s}}{}^{2}}\bigg)Vv.
\end{align}

From the vanishing of the determinant of the principal-symbol matrix of the differential operator in Eq.~\eqref{Eq:ContPhiphi}, we obtain the equation of the characteristic curves of the continuity equation~\eqref{Eq:ContPhiphi}~\cite{Sommerfeld1949}:
\begin{equation}
\label{Eq:Char}
\mathcal{A}dx^{4}+\mathcal{B}dx^{2}dy^{2}+\mathcal{C}dy^{4}=0,
\end{equation}
where the coefficients \maths{\mathcal{A}}, \maths{\mathcal{B}}, and \maths{\mathcal{C}} are expressed in terms of \maths{\sigma/\rho}, \maths{A}, \maths{B}, \maths{C}, and \maths{D} as
\begin{align}
\label{Eq:A}
\mathcal{A}&=1-\frac{\sigma^{2}}{\rho^{2}}>0, \\
\label{Eq:B}
\mathcal{B}&=A+B-\frac{\sigma}{\rho}(C+D), \\
\label{Eq:C}
\mathcal{C}&=AB-CD.
\end{align}
Solving the quartic equation~\eqref{Eq:Char}  provides the characteristic velocities \maths{dx/dy} of Eq.~\eqref{Eq:ContPhiphi}:
\begin{equation}
\label{Eq:dx/dy}
\bigg(\frac{dx}{dy}\bigg)^{2}=\frac{-\mathcal{B}\pm\sqrt{\mathcal{B}^{2}-4\mathcal{A}\mathcal{C}}}{2\mathcal{A}}=\frac{2\mathcal{C}}{-\mathcal{B}\pm\sqrt{\mathcal{B}^{2}-4\mathcal{A}\mathcal{C}}}.
\end{equation}
As explained above, Eq.~\eqref{Eq:ContPhiphi} is strongly elliptic---and the flow is therefore superfluid~\cite{Josserand1997, Josserand1999}---when \maths{dx/dy\in i\mathbb{R}}. This condition is satisfied if and only if all three of the following inequalities hold:
\begin{equation}
\label{Eq:Elliptic}
\mathcal{B}>0,\quad\mathcal{C}>0,\quad \mathcal{B}^{2}-4\mathcal{A}\mathcal{C}>0.
\end{equation}
Although these local conditions may seem obscure in their raw mathematical form, they are, in fact, fully equivalent to a local-density generalization of the Landau criterion, which holds true in the hydraulic approximation (see, e.g., Refs.~\cite{Pomeau1993a, Hakim1997, Huynh2024a, Huynh2024b} for single-component superfluids). This equivalence will be illustrated in Sec.~\ref{SubSec:Vc} in the case of an impenetrable obstacle potential.

A close inspection of Eqs.~\eqref{Eq:rhoHydrau},~\eqref{Eq:sigmaHydrau},~\eqref{Eq:Localcdcs},~\eqref{Eq:a}--\eqref{Eq:d}, and~\eqref{Eq:A}--\eqref{Eq:C} shows that inequalities~\eqref{Eq:Elliptic} involve only the mean and relative velocities in the regions where superfluidity breaks down [at \maths{(x,y)=(0,\pm w^{\pm})} or for \maths{r<w}]. To recast these conditions into a single constraint on the incident velocity \maths{V_{0}}, thereby identifying the critical speed \maths{V_{\mathrm{c}}}, we must first determine how these local velocities are related to \maths{V_{0}}. This is the purpose of the next section.

\subsection{Incompressible flow}
\label{SubSec:Incomp}

To determine the mean and relative velocity fields \maths{\mathbf{V}} and \maths{\mathbf{v}}, we treat the flow as \textit{incompressible}. This involves assuming that the total and relative densities \maths{\rho} and \maths{\sigma} are independent of velocity, which is equivalent to neglecting the kinetic terms in the Bernoulli equations~\eqref{Eq:rhoHydrau} and~\eqref{Eq:sigmaHydrau}. In this limit, \maths{\rho} and \maths{\sigma} become piecewise-constant functions of the radial coordinate \maths{r}:
\begin{align}
\label{Eq:rhoIncomp}
\frac{(1+\alpha)\rho}{1+\beta\rho}&=\mu_{0}-U\Theta(w-r), \\
\label{Eq:sigmaIncomp}
\frac{(1-\alpha)\sigma}{1+\beta\rho}&=-u\Theta(w-r),
\end{align}
and the continuity equations~\eqref{Eq:ContStat} reduce to a simple Laplace problem for the flow potentials \maths{\Phi} and \maths{\phi}:
\begin{equation}
\label{Eq:Laplace}
\nabla^{2}\!
\begin{bmatrix}
\Phi \\ \phi
\end{bmatrix}
\underset{r\gtrless w}{=}
\begin{bmatrix}
0 \\ 0
\end{bmatrix}
,\quad
\begin{bmatrix}
\Phi \\ \phi
\end{bmatrix}
\underset{r\gg w}{\simeq}
\begin{bmatrix}
2V_{0}r\cos\varphi \\ 0
\end{bmatrix}
,
\end{equation}
where \maths{\varphi} denotes the angular coordinate of \maths{\mathbf{r}=r(\hat{\mathbf{x}}\cos\varphi+\hat{\mathbf{y}}\sin\varphi)}. Note that this incompressible approach, first introduced in Ref.~\cite{Frisch1992}, was later extended by Rica~\cite{Rica2001} using a Janzen-Rayleigh expansion~\cite{Janzen1913, Rayleigh1916}, a standard technique for capturing compressibility effects. Formally, this involves an expansion of the velocity potentials in the compressibility \maths{1/(kc^{2})=1/(g\rho_{0})} relative to \maths{1/(kV_{0}{}^{2})}. Although the superfluid is not strictly incompressible, recent findings~\cite{Huynh2024a, Huynh2024b} indicate that the leading-order, incompressible approximation is sufficient to capture the essential features of the critical speed across a wide range of obstacle strengths. Given its demonstrated accuracy in the hydraulic regime, we adopt this simpler incompressible framework here.

From Eqs.~\eqref{Eq:rhoIncomp} and~\eqref{Eq:sigmaIncomp}, we now express the densities \maths{\rho} and \maths{\sigma} as functions of the obstacle amplitudes \maths{U} and \maths{u}. This step, essential for determining the velocity fields, also allows us to establish the domain of validity of the incompressible approximation in terms of \maths{U} and \maths{u}. Outside the obstacle, in the  region \maths{r>w} of the plane, \maths{\rho} and \maths{\sigma} are simply [use Eq.~\eqref{Eq:mu0} for \maths{\mu_{0}}]
\begin{equation}
\label{Eq:rhosigma_out}
\rho_{\mathrm{out}}=1,\quad\sigma_{\mathrm{out}}=0,
\end{equation}
while inside, for \maths{r<w}, they take the form
\begin{equation}
\label{Eq:rhosigma_in}
\rho_{\mathrm{in}}=\frac{1-U/\mu_{0}}{1+\beta U/\mu_{0}},\quad\sigma_{\mathrm{in}}=-\frac{1+\alpha}{1-\alpha}\frac{u/\mu_{0}}{1+\beta U/\mu_{0}}.
\end{equation}
For the first identity in~\eqref{Eq:rhosigma_in} to be physically meaningful, i.e., \maths{\rho_{\mathrm{in}}>0}, the total potential amplitude \maths{U} must satisfy \maths{U/\mu_{0}<1}. Together with the condition in Eqs.~\eqref{Eq:U} that the obstacle be repulsive, we thus obtain
\begin{equation}
\label{Eq:UCond}
0<\frac{U}{\mu_{0}}<1.
\end{equation}
We deduce that for \maths{U/\mu_{0}>1}, \maths{\rho_{\mathrm{in}}} must identically vanish since it cannot be physically negative. This imposes \maths{\rho_{+,\mathrm{in}}=\rho_{-,\mathrm{in}}=0}, and thus \maths{\sigma_{\mathrm{in}}} must also vanish. In this regime, the superfluid is completely excluded from the interior of the obstacle potential, which can therefore be described as \textit{impenetrable}. In the \textit{penetrable} regime~\eqref{Eq:UCond}, the condition \maths{|\sigma_{\mathrm{in}}|/\rho_{\mathrm{in}}=|\!\cos\theta_{\mathrm{in}}|<1} must additionally be satisfied for the second identity in~\eqref{Eq:rhosigma_in} to be well defined. This further constrains the relative potential amplitude \maths{u} to \maths{|u|/\mu_{0}<[(1-\alpha)/(1+\alpha)](1-U/\mu_{0})}. Given the repulsiveness condition in Eqs.~\eqref{Eq:u}, \maths{u} must therefore lie within the range
\begin{equation}
\label{Eq:uCond}
\frac{|u|}{\mu_{0}}<
\begin{dcases}
\frac{U}{\mu_{0}} & \text{if} \quad \frac{U}{\mu_{0}}<\frac{1-\alpha}{2} \\
\frac{1-\alpha}{1+\alpha}\bigg(1-\frac{U}{\mu_{0}}\bigg) & \text{if} \quad \frac{U}{\mu_{0}}>\frac{1-\alpha}{2}
\end{dcases}
.
\end{equation}
In summary, two distinct regimes emerge within the incompressible approximation for the velocity fields. When \maths{U/\mu_{0}>1}, the obstacle potential is impenetrable, excluding the superfluid from its interior, so that \maths{\rho_{\mathrm{in}}=\sigma_{\mathrm{in}}=0}. Conversely, in the penetrable regime~\eqref{Eq:UCond}, the superfluid penetrates the obstacle and the interior densities \maths{\rho_{\mathrm{in}}} and \maths{\sigma_{\mathrm{in}}} are given by Eqs.~\eqref{Eq:rhosigma_in}. Within this regime, the current theory is limited to the \maths{U{-}u} domain specified by~\eqref{Eq:uCond}. 

With this established, we now solve the Laplace problem~\eqref{Eq:Laplace}, which in the presence of the obstacle must be supplemented with the following continuity relations at the \maths{r=w} boundary:
\begin{align}
\label{Eq:BC1}
\begin{bmatrix}
\Phi \\ \phi
\end{bmatrix}
_{r=w^{+}}&=
\begin{bmatrix}
\Phi \\ \phi
\end{bmatrix}
_{r=w^{-}}, \\
\label{Eq:BC2}
\begin{bmatrix}
\partial_{r}\Phi \\ \partial_{r}\phi
\end{bmatrix}
_{r=w^{+}}&=\rho_{\mathrm{in}}
\begin{bmatrix}
\partial_{r}\Phi \\ \partial_{r}\phi
\end{bmatrix}
_{r=w^{-}}+\sigma_{\mathrm{in}}
\begin{bmatrix}
\partial_{r}\phi \\ \partial_{r}\Phi
\end{bmatrix}
_{r=w^{-}}.
\end{align}
While Eqs.~\eqref{Eq:BC1} ensure the continuity of the velocity potentials \maths{\Phi} and \maths{\phi}, Eqs.~\eqref{Eq:BC2} enforce the continuity of the radial component of the current densities \maths{\rho\mathbf{V}+\sigma\mathbf{v}/2} and \maths{\sigma\mathbf{V}+\rho\mathbf{v}/2} [we have used Eqs.~\eqref{Eq:rhosigma_out} for the exterior densities], which is a consequence of~\eqref{Eq:ContStat} and holds in the hydraulic limit \maths{w\gg1}~\cite{Huynh2024a, Huynh2024b}. We solve the Laplace equations in~\eqref{Eq:Laplace} using the generic solution \maths{a_{0}+b_{0}\ln r+\sum_{n\geqslant1}r_{n}(r)\varphi_{n}(\varphi)}, where \maths{r_{n}(r)=a_{n}r^{n}+b_{n}r^{-n}}, \maths{\varphi_{n}(\varphi)=c_{n}\cos(n\varphi)+d_{n}\sin(n\varphi)}, and the coefficients \maths{a_{0},\dots,d_{n}} are determined from the asymptotic and boundary conditions in Eqs.~\eqref{Eq:Laplace},~\eqref{Eq:BC1}, and~\eqref{Eq:BC2}.\footnote{We can also use the method of complex variables for planar potential flows~\cite{Batchelor2000}, as done, e.g., in Ref.~\cite{Rica2001}.} The velocity fields \maths{\mathbf{V}=\nabla\Phi/2} and \maths{\mathbf{v}=\nabla\phi} are found to be
\begin{align}
\label{Eq:V}
\mathbf{V}&=V_{0}
\begin{dcases}
\hat{\mathbf{x}}-\Lambda\frac{w^{2}}{r^{2}}(\hat{\mathbf{r}}\cos\varphi+\hat{\boldsymbol{\varphi}}\sin\varphi) & \text{if} \quad r>w \\
(1+\Lambda)\hat{\mathbf{x}} & \text{if} \quad r<w
\end{dcases}
, \\
\label{Eq:v}
\mathbf{v}&=V_{0}
\begin{dcases}
\lambda\frac{w^{2}}{r^{2}}(\hat{\mathbf{r}}\cos\varphi+\hat{\boldsymbol{\varphi}}\sin\varphi) & \text{if} \quad r>w \\
-\lambda\hat{\mathbf{x}} & \text{if} \quad r<w
\end{dcases}
,
\end{align}
where the scattering amplitudes
\begin{equation}
\label{Eq:Lambdalambda}
\Lambda=\frac{1-\rho_{\mathrm{in}}{}^{2}+\sigma_{\mathrm{in}}{}^{2}}{(1+\rho_{\mathrm{in}})^{2}-\sigma_{\mathrm{in}}{}^{2}},\quad\lambda=\frac{4\sigma_{\mathrm{in}}}{(1+\rho_{\mathrm{in}})^{2}-\sigma_{\mathrm{in}}{}^{2}}.
\end{equation}
As indicated in Sec.~\ref{SubSec:Elliptic}, the superfluid conditions~\eqref{Eq:Elliptic} are derived in the vicinity of the obstacle's poles, at \maths{(r,\varphi)=(w^{+},\pm\pi/2)}, or inside the obstacle, for \maths{r<w}. At these locations, Eqs.~\eqref{Eq:V} and~\eqref{Eq:v} yield \maths{\mathbf{V}=V_{0}(1+\Lambda)\hat{\mathbf{x}}} and \maths{\mathbf{v}=-V_{0}\lambda\hat{\mathbf{x}}} for the total and relative velocity, respectively. Notice that these are collinear to the mean asymptotic flow, thereby confirming the initial ansatz~\eqref{Eq:LocalVv}. Consequently, the magnitudes and signs of the local velocities appearing in~\eqref{Eq:Elliptic} must be identified to
\begin{equation}
\label{Eq:Vv}
\begin{dcases}
V=V_{0}|1+\Lambda| \\
s=\mathrm{sgn}(1+\Lambda)
\end{dcases}
,\quad
\begin{dcases}
v=V_{0}|\lambda| \\
s'=-\mathrm{sgn}(\lambda)
\end{dcases}
.
\end{equation}

In the next section, we simplify conditions~\eqref{Eq:Elliptic} into a single criterion for the mean asymptotic velocity \maths{V_{0}}. The analysis relies on Eqs.~\eqref{Eq:Localcdcs},~\eqref{Eq:a}--\eqref{Eq:d}, and~\eqref{Eq:A}--\eqref{Eq:C}, with the densities taken in the hydraulic approximation from Eqs.~\eqref{Eq:rhoHydrau} and~\eqref{Eq:sigmaHydrau}, and the velocities obtained in the incompressible approximation from Eqs.~\eqref{Eq:Lambdalambda} and~\eqref{Eq:Vv}. We first address the impenetrable-obstacle regime \maths{U/\mu_{0}>1} (Sec.~\ref{SubSubSec:VcImpene}), where \maths{\rho_{\mathrm{in}}=\sigma_{\mathrm{in}}=0}. We then turn to the penetrable-obstacle regime (Sec.~\ref{SubSubSec:VcPene}), where \maths{\rho_{\mathrm{in}}} and \maths{\sigma_{\mathrm{in}}} are given by Eqs.~\eqref{Eq:rhosigma_in} within the \maths{U{-}u} domain defined by~\eqref{Eq:UCond} and~\eqref{Eq:uCond}.

\subsection{Critical speed}
\label{SubSec:Vc}

\subsubsection{Impenetrable obstacle}
\label{SubSubSec:VcImpene}

When the obstacle potential is impenetrable, i.e., \maths{U/\mu_{0}>1}, superfluidity breaks down at the obstacle's poles \maths{(r,\varphi)=(w^{+},\pm\pi/2)}, as will be  shown numerically in Sec.~\ref{SubSec:Num}. At these points, the relative velocity \maths{v} given in Eqs.~\eqref{Eq:Lambdalambda} and~\eqref{Eq:Vv} is zero, because \maths{\lambda\propto\sigma_{\mathrm{in}}=0}. As a result, the relative density \maths{\sigma} deduced from Eq.~\eqref{Eq:sigmaHydrau} at \maths{r=w^{+}} also vanishes:
\begin{equation}
\label{Eq:sigmavImpene}
\sigma=0,\quad v=0.
\end{equation}
Equations~\eqref{Eq:a}--\eqref{Eq:d} therefore reduce to \maths{A=1-V^{2}/\tilde{c}_{\mathrm{d}}{}^{2}}, \maths{B=1-V^{2}/\tilde{c}_{\mathrm{s}}{}^{2}}, and \maths{C=D=0}, and Eqs.~\eqref{Eq:A}--\eqref{Eq:C} become \maths{\mathcal{A}=1}, \maths{\mathcal{B}=A+B}, and \maths{\mathcal{C}=AB}. This being so, the last superfluid constraint in~\eqref{Eq:Elliptic} provides no further information, as it is automatically satisfied: \maths{\mathcal{B}^{2}-4\mathcal{A}\mathcal{C}=(A-B)^{2}} is indeed always positive. The remaining constraints are \maths{A+B>0} and \maths{AB>0}, which yields the two inequalities
\begin{equation}
\label{Eq:LocalLandau}
V<\tilde{c}_{\mathrm{d}},\quad V<\tilde{c}_{\mathrm{s}}.
\end{equation}
These conditions represent a local formulation of the Landau criterion~\eqref{Eq:VcLandau}: for the flow to remain superfluid, the local mean velocity must be smaller than both the local density and spin speeds of sound. While these conditions are derived at the obstacle's poles following Refs.~\cite{Josserand1997, Josserand1999}, Chaplygin's method~\cite{Chaplygin1902, Landau1987} ensures their validity across the entire plane. Crucially, it can be shown from Eqs.~\eqref{Eq:rhoHydrau},~\eqref{Eq:Localcdcs}, and~\eqref{Eq:V} that \maths{V(\mathbf{r})} is maximized at the poles, whereas \maths{\tilde{c}_{\mathrm{d}}(\mathbf{r})} and \maths{\tilde{c}_{\mathrm{s}}(\mathbf{r})} are simultaneously minimized there. Consequently, satisfying the criterion at these specific points is both necessary and sufficient to ensure superfluidity globally, providing a rigorous justification for the polar analysis pioneered in Refs.~\cite{Josserand1997, Josserand1999}. This local Landau criterion is also physically consistent with the hydraulic approximation~\cite{Pomeau1993a, Hakim1997, Huynh2024a, Huynh2024b}, which assumes smoothly varying densities and thus supports a local-density description.

In inequalities~\eqref{Eq:LocalLandau}, the mean velocity \maths{V} is replaced with \maths{2V_{0}} using Eqs.~\eqref{Eq:Lambdalambda} and~\eqref{Eq:Vv} with \maths{\Lambda=1} (which follows from \maths{\rho_{\mathrm{in}}=\sigma_{\mathrm{in}}=0}). The total density \maths{\rho} in the local speeds of sound \maths{\tilde{c}_{\mathrm{d}}} and \maths{\tilde{c}_{\mathrm{s}}}, on the other hand, is deduced from Eq.~\eqref{Eq:rhoHydrau} at \maths{r=w^{+}}, setting \maths{V=2V_0}  and \maths{v=0} according to~\eqref{Eq:sigmavImpene}:
\begin{equation}
\label{Eq:VrhoImpene}
\rho=\frac{1-3V_{0}{}^{2}/\mu_{0}}{1+3\beta V_{0}{}^{2}/\mu_{0}},\quad V=2V_{0}.
\end{equation}
In turn, superfluid flow, encapsulated in inequalities~\eqref{Eq:LocalLandau}, is ensured as long as
\begin{equation}
\label{Eq:VcImpene}
V_{0}<V_{\mathrm{c}}=\min\{V_{1},V_{2}\},
\end{equation}
where the upper bounds \maths{V_{1}} [obtained from the first inequality in~\eqref{Eq:LocalLandau}] and \maths{V_{2}} (second inequality) are expressed in units of the Landau speeds of sound \maths{c_{\mathrm{d}}} and \maths{c_{\mathrm{s}}}, respectively, as
\begin{align}
\label{Eq:VdImpene}
\frac{V_{1}}{c_{\mathrm{d}}}&=\sqrt{\frac{(1+\beta)[\sqrt{(1+\beta)(121+25\beta)}-(11+5\beta)]}{9\beta}}, \\
\label{Eq:VsImpene}
\frac{V_{2}}{c_{\mathrm{s}}}&=\sqrt{\frac{2(1+\alpha)}{11+5\alpha}}.
\end{align}
Condition~\eqref{Eq:VcImpene} defines  \maths{V_{\mathrm{c}}} as the  critical speed for superfluidity in the impenetrable-obstacle regime, within the hydraulic and incompressible approximations. Based on the mathematical decoupling of the ellipticity conditions~\eqref{Eq:Elliptic} into the mode-specific constraints \maths{V<\tilde{c}_{\mathrm{d}}} and \maths{V<\tilde{c}_{\mathrm{s}}} [arising from the vanishing of the relative fields in Eqs.~\eqref{Eq:sigmavImpene}], one might naively associate \maths{V_{1}} and \maths{V_{2}} with the critical speeds in the density and spin sectors, respectively. However, as will be shown in Sec.~\ref{SubSec:Num}, the physical excitations generated above \maths{V_{\mathrm{c}}=\min\{V_{1},V_{2}\}} appear indistinguishably in both the density and spin maps. Indeed, in the impenetrable regime, the obstacle naturally engenders a strongly nonlinear response that typically results in significant mode hybridization. This physical complexity motivates our use of the \maths{V_{1(2)}} notation, purposely avoiding a more reductive labeling such as \maths{V_{\mathrm{d}(\mathrm{s})}}.

\begin{figure}[t!]
\begin{center}
\includegraphics[width=\linewidth]{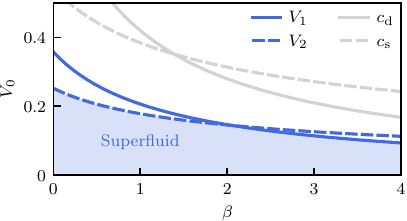}
\end{center}
\caption{Threshold speeds \maths{V_{1}} [Eq.~\eqref{Eq:VdImpene}; dark solid curve] and \maths{V_{2}} [Eq.~\eqref{Eq:VsImpene}; dark dashed curve] for an impenetrable obstacle potential (\maths{U/\mu_{0}>1}). While \maths{V_{1}} and \maths{V_{2}} are formally derived from the density- and spin-sector local constraints~\eqref{Eq:LocalLandau}, respectively, the resulting excitations above \maths{V_{\mathrm{c}}=\min\{V_{1},V_{2}\}} are hybridized (see Sec.~\ref{SubSec:Num}). The curves are plotted as functions of the saturation parameter \maths{\beta} for the intercomponent interaction constant \maths{\alpha\simeq0.41} of Ref.~\cite{Piekarski2025}. The light curves correspond to the linear-response, density and spin speeds of sound \maths{c_{\mathrm{d}}} and \maths{c_{\mathrm{s}}}, originally shown in Fig.~\ref{Fig:Landau}. The shaded area highlights the parameter regime where inequality~\eqref{Eq:VcImpene} holds, and the flow of mean asymptotic velocity \maths{V_{0}} is superfluid. Below (above) the intersection point \maths{\beta=\alpha(11+5\alpha)/(4-3\alpha-\alpha^{2})\simeq2.06}, the critical speed for superfluidity is \maths{V_{\mathrm{c}}=V_{2}} (\maths{V_{\mathrm{c}}=V_{1}}). In the Kerr regime (\maths{\beta\ll1}), \maths{V_{\mathrm{c}}=V_{2}\simeq0.25} while \maths{V_{1}\simeq0.36}.}
\label{Fig:Impenetrable}
\end{figure}

As seen from Eqs.~\eqref{Eq:VdImpene} and~\eqref{Eq:VsImpene}, \maths{V_{1}} and \maths{V_{2}} do not depend on the obstacle's amplitudes \maths{U} and \maths{u}, but only on the superfluid's parameters \maths{\alpha} and \maths{\beta}. In Fig.~\ref{Fig:Impenetrable}, we plot them as functions of the saturation parameter \maths{\beta}, using the interaction constant \maths{\alpha\simeq0.41} from Ref.~\cite{Piekarski2025}, as in Figs.~\ref{Fig:Landau} and~\ref{Fig:Drag}. 
We observe that they are essentially a renormalization of the Landau critical speeds \maths{c_{\mathrm{d}}} and \maths{c_{\mathrm{s}}}, shown in light color for comparison. They exhibit a similar trend as a function of \maths{\beta} and, in particular, display a similar inversion occurring at \maths{\beta=\alpha(11+5\alpha)/(4-3\alpha-\alpha^{2})}. Below (above) this threshold, the critical speed for superfluidity is \maths{V_{\mathrm{c}}=V_{2}} (\maths{V_{\mathrm{c}}=V_{1}}) according to condition~\eqref{Eq:VcImpene}, which is fulfilled in the shaded region of the \maths{\beta{-}V_{0}} plane. An interesting limit is the Kerr regime  \maths{\beta\ll1}, where one has
\begin{equation}
\label{Eq:VdVsKerr}
V_{1}\simeq\sqrt{\frac{2}{11}}c_{\mathrm{d}},\quad V_{\mathrm{c}}=V_{2}=\sqrt{\frac{2(1+\alpha)}{11+5\alpha}}c_{\mathrm{s}},
\end{equation}
with \maths{c_{\mathrm{d}}\simeq[(1+\alpha)/2]^{1/2}} and \maths{c_{\mathrm{s}}\simeq[(1-\alpha)/2]^{1/2}} (see Sec.~\ref{SubSec:Landau}). In this regime and for a single-component superfluid (\maths{\alpha=0}), the critical speed \maths{V_{\mathrm{c}}=V_{2}=V_{1}=(2/11)^{1/2}/\sqrt{2}} matches the well-known result for a 2D scalar cubic nonlinear Schr\"odinger superflow~\cite{Frisch1992, Rica2001} with background speed of sound \maths{[(1\pm\alpha)/2]^{1/2}=1/\sqrt{2}}.

\subsubsection{Penetrable obstacle}
\label{SubSubSec:VcPene}

For a repulsive and penetrable obstacle potential [condition~\eqref{Eq:UCond}], superfluidity breaks down inside the obstacle, as will be  shown numerically in Sec.~\ref{SubSec:Num}. In this region of the plane, the densities within the hydraulic approximation are given by Eqs.~\eqref{Eq:rhoHydrau} and~\eqref{Eq:sigmaHydrau} for \maths{r<w}, while the velocities in the incompressible approximation are obtained from Eqs.~\eqref{Eq:Lambdalambda} and~\eqref{Eq:Vv}, using the constant interior densities \maths{\rho_{\mathrm{in}}} and \maths{\sigma_{\mathrm{in}}} defined in Eqs.~\eqref{Eq:rhosigma_in} for \maths{u} in the range~\eqref{Eq:uCond}.

In this regime, the three superfluidity conditions~\eqref{Eq:Elliptic} also reduce to a local Landau criterion, though not as straightforward as criterion~\eqref{Eq:LocalLandau} for an impenetrable obstacle. This criterion is a very nontrivial combination of local-density extensions of the two Landau criteria established in Appendix~\ref{App:Imbalanced}. This complexity arises from the fact that the relative potential amplitude \maths{u} is generically nonzero, which yields both nonvanishing relative fields \maths{\sigma} and \maths{v} in the core equations~\eqref{Eq:a}--\eqref{Eq:d} and~\eqref{Eq:A}--\eqref{Eq:C}. Nevertheless, a symbolic-calculus analytical treatment of this local Landau criterion remains possible and eventually leads to a condition of the form
\begin{equation}
\label{Eq:VcPene}
V_{0}<V_{\mathrm{c}}(U,|u|,\alpha,\beta),
\end{equation}
where the critical speed \maths{V_{\mathrm{c}}} now depends explicitly on the obstacle strengths \maths{U} and \maths{u}, unlike in Eqs.~\eqref{Eq:VcImpene}--\eqref{Eq:VsImpene}. Interestingly, its dependence on the relative amplitude \maths{u} is solely on its absolute value, \maths{|u|}. We do not provide its exact expression here, as it exhibits prohibitively cumbersome dependencies on \maths{U}, \maths{|u|}, \maths{\alpha}, and \maths{\beta}.

\begin{figure}[t!]
\begin{center}
\includegraphics[width=\linewidth]{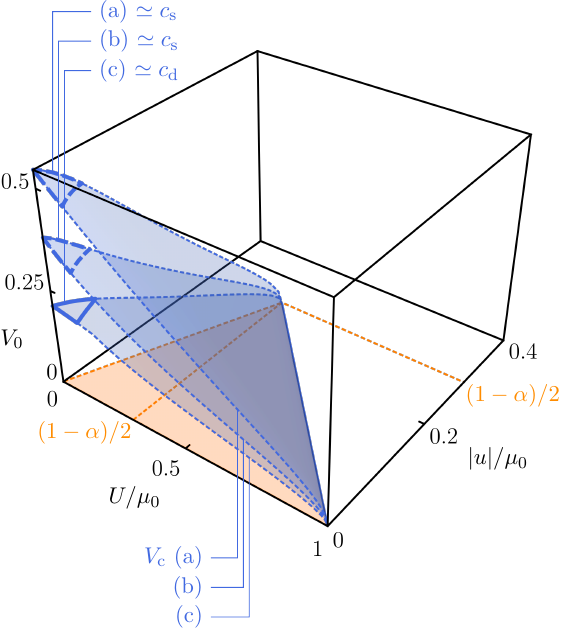}
\end{center}
\caption{Surface plot of the critical speed \maths{V_{\mathrm{c}}(U,|u|,\alpha,\beta)} as a function of the normalized obstacle amplitudes \maths{U/\mu_{0}} and \maths{|u|/\mu_{0}} in the penetrable regime~\eqref{Eq:UCond}. The orange region in the \maths{U/\mu_{0}{-}|u|/\mu_{0}} plane corresponds to the range of applicability~\eqref{Eq:uCond} of our approach.  In this plot, the interaction constant is set to the value \maths{\alpha\simeq0.41} used in the experiment of Ref.~\cite{Piekarski2025}. The blue, shaded sheets show \maths{V_{\mathrm{c}}} for (a)~\maths{\beta=0} (Kerr regime), (b)~\maths{\beta=1}, and (c)~\maths{\beta=3}. For \maths{U/\mu_{0},|u|/\mu_{0}\ll1}, \maths{V_{\mathrm{c}}} converges to the linear-response predictions: (a)~\maths{c_{\mathrm{s}}\simeq0.54}, (b)~\maths{c_{\mathrm{s}}\simeq0.38}, and (c)~\maths{c_{\mathrm{d}}\simeq0.21} (see Fig.~\ref{Fig:Landau}).}
\label{Fig:Penetrable}
\end{figure}

Instead, Fig.~\ref{Fig:Penetrable} graphically shows \maths{V_{\mathrm{c}}} as a function of \maths{U/\mu_{0}} and \maths{|u|/\mu_{0}}. This critical surface (represented as a blue, shaded sheet) is plotted for the interaction constant \maths{\alpha\simeq0.41} of experiment~\cite{Piekarski2025} and three saturation parameters: \maths{\beta=0}, corresponding to the Kerr regime, and \maths{\beta=1} and \maths{\beta=3}. These values of \maths{\beta} are chosen to lie on either side of the transition point observed in the weak- and impenetrable-obstacle regimes---where, in the former, the critical speed shifts from the spin to the density sector, while in the latter, the limiting threshold switches between \maths{V_{2}} and \maths{V_{1}} (see Fig.~\ref{Fig:Impenetrable}). The critical speed is defined within a triangular support corresponding to the range of applicability~\eqref{Eq:uCond} of our approach. Within this range, the parameter region where the flow remains superfluid lies below the critical surface. The figure shows that the critical speed   \maths{V_{\mathrm{c}}} for a penetrable obstacle decreases with the potential amplitudes \maths{U} and \maths{|u|}. This is consistent with the physical intuition that a stronger perturbation is more detrimental to superfluidity. Furthermore, for fixed values of \maths{U} and \maths{|u|}, the critical speed is higher at smaller saturation \maths{\beta}, a behavior also observed for weak and impenetrable obstacles (see Fig.~\ref{Fig:Impenetrable}).

The boundaries of the \maths{V_\mathrm{c}} surface also deserve a few comments. First, in the  region \maths{U/\mu_{0},|u|/\mu_{0}\ll1} where linear-response theory is valid, we recover that \maths{V_\mathrm{c}} converges to the Landau critical speed \maths{c_\mathrm{s}\simeq0.54} (\maths{c_\mathrm{s}\simeq0.38}, \maths{c_\mathrm{d}\simeq0.21}) when \maths{\beta=0} (\maths{\beta=1}, \maths{\beta=3}), in agreement with Fig.~\ref{Fig:Landau}. Second, near the boundary \maths{u=0}, i.e., for a nearly scalar obstacle, the critical speed takes a relatively simple form in the Kerr limit \maths{\beta\ll1}:
\begin{align}
\label{Eq:VdKerru0}
\notag
&\left.V_{\mathrm{c}}(U,|u|\ll\mu_{0},\alpha,\beta\ll1)\right. \\
&\left.\quad\simeq\sqrt{\frac{2(1+\alpha)(1-\tilde{U})(2-\tilde{U})^{2}}{8+\tilde{U}(4-\tilde{U})+\alpha[8-\tilde{U}(4-\tilde{U})]}}c_{\mathrm{s}},\right.
\end{align}
where \maths{\tilde{U}=U/\mu_{0}} and \maths{c_{\mathrm{s}}\simeq[(1-\alpha)/2]^{1/2}} is the Kerr speed of sound in the spin mode. Note that in the weak-obstacle regime, the Landau critical speed is again recovered: \maths{V_{\mathrm{c}}(U\ll\mu_{0},|u|\ll\mu_{0},\alpha,\beta\ll1)\simeq c_{\mathrm{s}}}. Third, Fig.~\ref{Fig:Penetrable} shows that \maths{V_{\mathrm{c}}\to0} when \maths{(U/\mu_{0},|u|/\mu_{0})\to(1,0)}, suggesting an apparent discontinuity with the finite value of \maths{V_{\mathrm{c}}} found in Sec.~\ref{SubSubSec:VcImpene} for the impenetrable case \maths{U/\mu_{0}>1}. In fact, the boundary between the penetrable and impenetrable regimes, while sharp at the level of our approximations, is expected to be smoothed by quantum-pressure effects, as shown in Refs.~\cite{Kwak2023, Huynh2024a, Huynh2024b}. Finally, it is worthwhile to stress that our results are restricted to the orange region of obstacle amplitudes in Fig.~\ref{Fig:Penetrable}. Outside this region, the present framework must be revisited: see Sec.~\ref{Sec:Outro} for further discussion about this point.

In the next section, we complement our analytical results with numerical simulations of the full wave equations in~\eqref{Eq:dNLS}. This serves a dual purpose: first, to provide physical insight into the excitations responsible for the breakdown of superfluidity, and second, to provide validation of our analytical predictions for specific obstacle and fluid parameters. Furthermore, these simulations allow us to test the hypothesis that these excitations are generated either at the poles of the obstacle or within its core.

\subsection{Numerical simulations }
\label{SubSec:Num}

To gain deeper insight into the nature of the excitations accompanying the breakdown of superfluidity in the nonlinear regime, we complement the theoretical analysis presented above with numerical simulations whose details are provided in Appendix~\ref{App:Num}. We consider the real-time propagation of a 2D quasi-plane-wave beam with two balanced components past a two-amplitude circular potential barrier of radius \maths{w} with a smooth hyperbolic-tangent boundary of small width \maths{\delta w} [see Eq.~\eqref{Eq:UpmNum}]. The propagation along the \maths{z} axis, ruled by the two coupled equations in~\eqref{Eq:dNLS}, is simulated using the standard split-step Fourier method~\cite{Agrawal2019}. We fix the obstacle geometry to \maths{w=10} and \maths{\delta w=1}, the interaction parameter to \maths{\alpha=0.41} (consistent with experiment~\cite{Piekarski2025}), and the saturation parameter to \maths{\beta=0} (corresponding to the Kerr regime) without loss of generality (\maths{\beta\neq0} leads to a similar phenomenology for the numerically observed excitations). The remaining control parameters in the simulations are the total and relative obstacle amplitudes, \maths{U/\mu_{0}} and \maths{u/\mu_{0}} with \maths{\mu_{0}} from~\eqref{Eq:mu0}, and the input mean flow velocity \maths{V_{0}} in the positive-\maths{x} direction.

\begin{figure}[t!]
\begin{center}
\includegraphics[width=\linewidth]{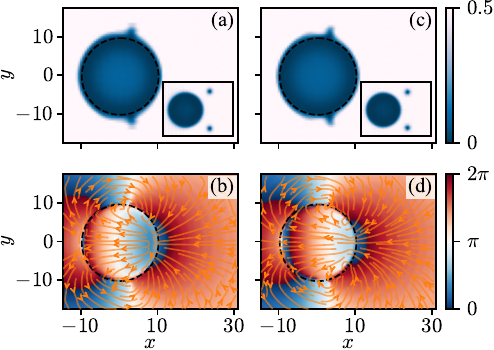}
\end{center}
\caption{Evolution along the propagation axis---\maths{z=80} for the main images; \maths{z=200} for the insets---of the 2D transverse distributions for the two superfluid components (a,~b)~\maths{\psi_{+}} and (c,~d)~\maths{\psi_{-}} in the impenetrable-obstacle regime (\maths{U/\mu_{0}>1}). Panels~(a) and~(c) display the densities \maths{\rho_{+}(x,y)} and \maths{\rho_{-}(x,y)}, respectively, while~(b) and~(d) show the corresponding phase maps \maths{\phi_{+}(x,y)} and \maths{\phi_{-}(x,y)} (color scale) and velocity fields \maths{\nabla\phi_{+}(x,y)\equiv\mathbf{v}_{+}(x,y)} and \maths{\nabla\phi_{-}(x,y)\equiv\mathbf{v}_{-}(x,y)} (orange streamlines). These snapshots are obtained from real-time numerical simulations of the coupled system~\eqref{Eq:dNLS}. The dashed circle indicates the edge of the obstacle. The densities reveal the formation of a pair of localized depletions in both components near the poles of the obstacle when the mean flow velocity exceeds the critical speed. At larger \maths{z}, these defects are shed and migrate downstream, as shown in the insets. The phase and velocity maps confirm the quantized-vortex nature of these excitations, characterized by opposite phase windings of \maths{+2\pi} (lower pole of the obstacle) and \maths{-2\pi} (upper pole). The obstacle radius (boundary width) is \maths{w=10} (\maths{\delta w=1}), and its total and relative amplitudes are \maths{U/\mu_{0}=8} and \maths{|u|/\mu_{0}=2}. The flow parameters are \maths{\alpha=0.41}~\cite{Piekarski2025}, \maths{\beta=0} (Kerr limit), and \maths{V_{0}=0.29>V_{\mathrm{c}}\simeq0.27} (\maths{V_{\mathrm{c}}\simeq0.25} analytically).}
\label{Fig:ImpenetrableNum}
\end{figure}

We first focus on the impenetrable-obstacle regime (\maths{U/\mu_{0}>1}) studied in Sec.~\ref{SubSubSec:VcImpene}, choosing \maths{U/\mu_{0}=8} and \maths{|u|/\mu_{0}=2} as representative obstacle amplitudes. Figure~\ref{Fig:ImpenetrableNum} shows the corresponding densities, phases, and streamlines for the superfluid components (a, b)~\maths{\psi_{+}} and (c, d)~\maths{\psi_{-}} at specific effective propagation times \maths{z}. Panels~(a) and~(b) display \maths{\rho_{+}(x,y)} and \maths{[\phi_{+}(x,y),\nabla\phi_{+}(x,y)\equiv\mathbf{v}_{+}(x,y)]}, respectively, while panels~(c) and~(d) show the same fields for the \maths{-} wave. Note that in the present nonlinear regime, the flow dynamics exhibits an expected density-spin hybridization. We choose to represent its excitations via its individual \maths{+} and \maths{-} components, as density and spin representations tend to merge these features into complex, nontrivial structures that obscure the underlying physics. For the chosen values of \maths{U} and \maths{u}, we numerically estimate a critical velocity \maths{V_{\mathrm{c}}\simeq0.27}, which is in very good agreement with, albeit slightly higher than, the analytical prediction of Sec.~\ref{SubSubSec:VcImpene}. All the snapshots in Fig.~\ref{Fig:ImpenetrableNum} are obtained for a mean flow velocity \maths{V_{0}=0.29}, i.e., just above \maths{V_{\mathrm{c}}}. They clearly show the nucleation, in both the \maths{+} and \maths{-} components, of a quantized vortex (\maths{+2\pi} phase winding) near the lower pole of the obstacle, accompanied by an antivortex (\maths{-2\pi} phase winding) near the upper pole, before both are shed into the wake as illustrated in the insets of panels~(a) and~(c). Note that while the local velocity maximum is theoretically expected exactly at the poles (see Sec.~\ref{SubSubSec:VcImpene}), our numerical simulations---which account for both density gradients and compressibility---show that the effective shedding occurs slightly downstream. Indeed, the shedding dynamics favors the near-wake regions, where the local pressure and flow conditions can effectively ``tear'' the vortices away from the obstacle surface. We also note that the dispersive terms neglected in the hydraulic approximation---but fully accounted for in the numerical integration of~\eqref{Eq:dNLS}---allow for small evanescent penetrations of the wave functions into the high-potential region. Although their densities are here two orders of magnitude lower than in the exterior, their phases remain well-defined within the obstacle. This ensures the continuity of the corresponding velocity fields and explains why the streamlines in panels~(b) and~(d) persist into the interior of the high potential barrier.

\begin{figure}[t!]
\begin{center}
\includegraphics[width=\linewidth]{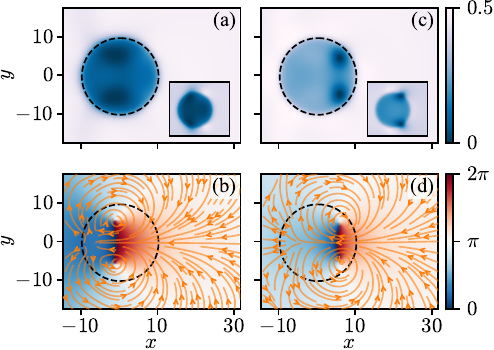}
\end{center}
\caption{Same as Fig.~\ref{Fig:ImpenetrableNum} but in the penetrable-obstacle regime: \maths{U/\mu_{0}<1} and condition~\eqref{Eq:uCond}; here \maths{U/\mu_{0}=0.6475} and \maths{|u|/\mu_{0}=0.07375} (see footnote~\ref{Foot:t}), with an effective propagation time \maths{z=80} for the main images and \maths{z=120} for the insets. In each component, (a,~b)~\maths{\psi_{+}} and (c,~d)~\maths{\psi_{-}}, a Jones-Roberts soliton forms inside the obstacle when the mean flow velocity exceeds the critical speed: here, \maths{V_{0}=0.15>V_{\mathrm{c}}\simeq0.13} (\maths{V_{\mathrm{c}}\simeq0.14} analytically). In the present configuration, panels~(b) and~(d) reveal closely spaced \maths{+2\pi} and \maths{-2\pi} phase windings, indicating that these Jones-Roberts solitons take the form of vortex-antivortex dipoles. The insets in panels~(a) and~(c) show their evolution at a later \maths{z}, highlighting trapping within the obstacle and dynamical shedding, respectively.}
\label{Fig:PenetrableNum}
\end{figure}

\begin{figure}[t!]
\begin{center}
\includegraphics[width=\linewidth]{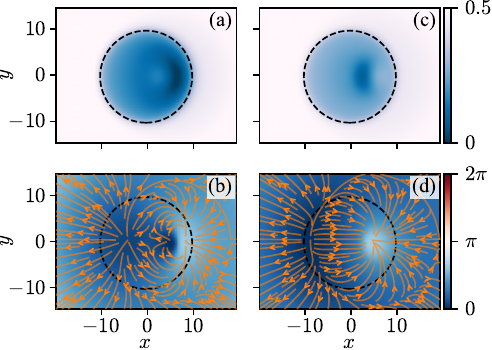}
\end{center}
\caption{Same as Fig.~\ref{Fig:PenetrableNum} but for \maths{U/\mu_{0}=0.45}, \maths{|u|/\mu_{0}=0.04}, \maths{V_{0}=0.15}, and \maths{z=30}. For this choice of parameters, the Jones-Roberts solitons nucleated inside the penetrable obstacle take the form of rarefaction waves, rather than vortex-antivortex dipoles.}
\label{Fig:RarefactionWave}
\end{figure}

Next, in Fig.~\ref{Fig:PenetrableNum}, we present the same hydrodynamic fields as in Fig.~\ref{Fig:ImpenetrableNum} but in the penetrable-obstacle regime [\maths{U/\mu_{0}<1} and condition~\eqref{Eq:uCond}] studied in Sec.~\ref{SubSubSec:VcPene}. Choosing \maths{U/\mu_{0}=0.6476} and \maths{|u|/\mu_{0}=0.0737},\footnote{\label{Foot:t}These obstacle amplitudes correspond to the point \maths{(U/\mu_{0},|u|/\mu_{0})=[(1+t)/2,t/4]} within the orange triangular support shown in Fig.~\ref{Fig:Penetrable}, with \maths{t\equiv(1-\alpha)/2=0.295} in the specific case where \maths{\alpha=0.41}.} we numerically estimate a critical velocity \maths{V_{\mathrm{c}}\simeq0.13}, in very good agreement with the analytical prediction \maths{V_{\mathrm{c}}\simeq0.14} [blue surface~(a) in Fig.~\ref{Fig:Penetrable}]. The snapshots correspond to a fluid velocity \maths{V_{0}=0.15}, i.e., slightly above \maths{V_{\mathrm{c}}}. They reveal that superfluidity breaks down through the formation of localized excitations inside the obstacle, as anticipated in Sec.~\ref{SubSubSec:VcPene}. These appear at different positions in panels~(a,~b) and~(c,~d), as a result of the different coupling of the two-amplitude obstacle potential to \maths{\psi_{+}} and \maths{\psi_{-}}, which manifests through asymmetric modifications of the local velocities in each component.\footnote{In the case of a scalar obstacle, \maths{u/\mu_{0}\to0} (\maths{U_{+}\simeq U_{-}}), these excitations are nucleated at the same position.} We confidently conjecture that these excitations are Jones-Roberts solitons~\cite{Jones1982, Meyer2017, BakerRasooli2025}, similar to those presumably observed in single-component superfluids of light~\cite{Eloy2021, Huynh2024b}. For the obstacle parameters considered in Fig.~\ref{Fig:PenetrableNum}, the phase and velocity maps in panels~(b) and~(d) exhibit closely spaced \maths{+2\pi} and \maths{-2\pi} phase windings. This indicates that these Jones-Roberts solitons effectively behave as vortex-antivortex dipoles. As shown in the insets of panels~(a) and~(c), the evolution at later propagation times \maths{z} reveals complex nonlinear dynamics. Depending on the component and the specific parameters, the structures may remain trapped within the obstacle, as seen in~(a), or migrate towards its edges to eventually detach and be carried downstream, as in~(c). This behavior highlights a competition between trapping within the potential and a shedding process analogous to that observed in the insets of Figs.~\ref{Fig:ImpenetrableNum}(a,~c). Importantly, we note that the topological nature of these excitations is highly sensitive to the specific choice of parameters. Indeed, for other obstacle amplitudes \maths{U} and \maths{u}, these Jones-Roberts solitons can instead lie in the rarefaction-wave regime of their dispersion relations. In this case, the phase singularities disappear, leaving single density depletions characterized by smooth phase variations across the structures, as illustrated in Fig.~\ref{Fig:RarefactionWave} where \maths{U/\mu_{0}=0.45}, \maths{|u|/\mu_{0}=0.04}, and \maths{V_{0}=\text{0.15}}.

\section{Conclusion and outlook}
\label{Sec:Outro}

We have studied the critical speed for dissipationless motion of a 2D binary superfluid of light flowing past a repulsive obstacle with polarization-dependent amplitudes. We have considered an optical fluid with saturable intra- and intercomponent interactions, focusing on the simplest case of a balanced mixture in the miscible regime.

The critical speed \maths{V_{\mathrm{c}}} corresponds to the lowest mean velocity below which no excitation is produced in the fluid. We have derived this speed analytically for both weak and arbitrary obstacle strengths. In the weak-obstacle regime, we have employed linear-response theory and showed that \maths{V_{\mathrm{c}}=\min\{c_{\mathrm{d}},c_{\mathrm{s}}\}}, in agreement with Landau's criterion, where \maths{c_{\mathrm{d}}} and \maths{c_{\mathrm{s}}} are the speeds of sound for the density and spin Bogoliubov modes. Their relative ordering depends on the degree of interaction saturation. These results are consistent with the behavior of the drag force experienced by the obstacle, which vanishes below \maths{V_{\mathrm{c}}} and rises sharply above \maths{c_{\mathrm{d}}} and \maths{c_{\mathrm{s}}}, signaling energy dissipation via the emission of density and spin Bogoliubov waves.

For arbitrary obstacle strengths, we have simplified the problem using the hydraulic approximation, assuming smooth density profiles, together with an incompressible approximation, where densities are taken to be velocity independent for the purpose of computing the flow. In this framework, the critical speed \maths{V_{\mathrm{c}}} is determined by the conditions for strong ellipticity of the two stationary continuity equations. It is lower than the Landau sound velocities and depends on the total and relative obstacle amplitudes \maths{U} and \maths{u}. We have calculated \maths{V_{\mathrm{c}}} for both penetrable and impenetrable obstacles. In the penetrable regime, it is a highly nontrivial decreasing function of \maths{U} and \maths{u}, while it saturates to a constant value in the impenetrable limit. We have confirmed the analytical estimates for \maths{V_{\mathrm{c}}} by real-time numerical simulations of the two coupled nonlinear Schr\"odinger equations. The simulations further show that superfluidity breaks down through the emission of vortex-antivortex pairs or Jones-Roberts solitons in both components, generated respectively at the poles or inside the obstacle for the impenetrable and penetrable cases.

The theoretical results presented in this paper naturally call for further investigation. First, for a penetrable obstacle, we have been able to derive the critical speed only within a region of obstacle amplitudes defined by condition~\eqref{Eq:uCond}. Outside this region, for \maths{[(1-\alpha)/(1+\alpha)](1-U/\mu_{0})<|u|/\mu_{0}<U/\mu_{0}}, it seems natural to assume, by continuity, that the ratio \maths{|\sigma_{\mathrm{in}}|/\rho_{\mathrm{in}}} saturates at its maximum value of unity. However, this assumption leads to the counterintuitive prediction of critical speeds exceeding the Landau speeds of sound. This issue could be clarified by incorporating dispersive effects into the theoretical framework~\cite{Huepe2000, Kokubo2025, Kanjo2026}, and by treating the superfluid compressibility more rigorously through Janzen-Rayleigh expansions, a task we leave for future work. Second, a dedicated analytical study of the magnetic, spin-type vortices and solitons observed in our numerical simulations is warranted. These still poorly studied nonlinear excitations are likely central to a complete understanding of dissipation mechanisms in 2D binary superfluids. Finally, while our analysis for repulsive interactions is restricted to the miscible regime (\maths{\alpha<1}), the immiscible case (\maths{\alpha>1}) represents a compelling direction for future studies. In this regime, any homogeneously mixed preparation is modulationally unstable in the spin channel, leading to spatial demixing of the two components~\cite{Pitaevskii2016}. Properly describing the onset of superfluidity would thus require accounting for a spatially heterogeneous background (characteristic of the immiscible ground state), a task that lies beyond the current analytical framework but promises rich physics regarding the interplay between phase separation and topological-defect nucleation.

\begin{acknowledgement}
We acknowledge Thorsten Ackemann, Mathias Albert, Quentin Glorieux, Juliette Huynh, Nicolas Pavloff, and Clara Piekarski for valuable discussions.
\end{acknowledgement}

\section*{Author contribution statement}

PEL developed the theoretical framework, performed the analytical calculations, prepared the corresponding figures, and wrote the manuscript. CM conducted the numerical simulations and prepared the corresponding figures. NC contributed to the writing and critical review of the manuscript and calculations.

\section*{Data availability statement}

Data sets generated during the current study are available from the corresponding author on reasonable request.

\section*{Declarations}

\begin{paragraph}{Funding}
This work was supported by the French National Research Agency (ANR) through grants ANR-21-CE47-0009 (Quantum-SOPHA), ANR-21-CE30-0008 (STLight), ANR-24-CE47-4949-03 (UniQ-RingS), and ANR-24-CE30-6695 (FUSIoN).
\end{paragraph}

\begin{paragraph}{Competing interests}
The authors declare no competing interests.
\end{paragraph}

\appendix
\numberwithin{equation}{section}

\begin{strip}

\section{Landau's criteria for imbalanced mixtures}
\label{App:Imbalanced}

This appendix provides results for Landau's critical speed when the incoming flow is imbalanced in density and velocity. For readability, we distinguish two situations: first, when \maths{\sigma_{0}=0} and \maths{\mathbf{v}_{0}\neq0}, which corresponds to an incoming flow balanced in density but imbalanced in velocity (Sec.~\ref{SubApp:ImbalancedVelocity}); and second, when \maths{\sigma_{0}\neq0} and \maths{\mathbf{v}_{0}=0}, representing an incoming flow imbalanced in density but balanced in velocity (Sec.~\ref{SubApp:ImbalancedDensity}).

\subsection{\maths{\boldsymbol{\sigma_{0}=0}} and \maths{\boldsymbol{\mathbf{v}_{0}\neq0}}}
\label{SubApp:ImbalancedVelocity}

In this configuration, the relative-velocity potential at infinity \maths{\phi\simeq\mathbf{v}_{0}\cdot\mathbf{r}} (up to an irrelevant additive constant), the chemical potential \maths{\mu_{\pm}=[\mu_{0}+(\mathbf{V}_{0}\pm\mathbf{v}_{0}/2)^{2}]/2}, where \maths{\mu_{0}} is defined in Eq.~\eqref{Eq:mu0}, and the matrix \maths{\mathbb{L}} in Eq.~\eqref{Eq:BdG} is given by
\begin{equation}
\label{Eq:Lv}
\mathbb{L}=
\begin{bmatrix}
0 & \frac{K^{2}}{2} & \frac{iv_{0}K\cos\alpha_{\mathbf{K}}}{2} & 0 \\
-\frac{K^{2}}{2}-2c_{\mathrm{d}}{}^{2} & 0 & 0 & -\frac{iv_{0}K\cos\alpha_{\mathbf{K}}}{2} \\
\frac{iv_{0}K\cos\alpha_{\mathbf{K}}}{2} & 0 & 0 & -\frac{K^{2}}{2} \\
0 & -\frac{iv_{0}K\cos\alpha_{\mathbf{K}}}{2} & \frac{K^{2}}{2}+2c_{\mathrm{s}}{}^{2} & 0
\end{bmatrix}
,
\end{equation}
where \maths{\alpha_{\mathbf{K}}=\angle(\mathbf{K},\mathbf{v}_{0})} and \maths{c_{\mathrm{d}}} and \maths{c_{\mathrm{s}}} are defined in Eqs.~\eqref{Eq:cdcs}. The nonzero \maths{\mathbf{v}_{0}}-entries of~\eqref{Eq:Lv} signal that the total and relative fluid's fluctuations get hybridized compared to Sec.~\ref{SubSec:Landau}, which in principle precludes any distinction between a density mode and a spin mode. Their dispersion relations even consist of two distinct branches, which we hereafter label ``\maths{1}'' and ``\maths{2}.'' They are given by \maths{\Omega_{0}=\pm\Omega_{1}} and \maths{\Omega_{0}=\pm\Omega_{2}}, where
\begin{equation}
\label{Eq:Omega12v}
\Omega_{1,2}{}=\sqrt{\frac{K^{2}}{2}\bigg\{\frac{K^{2}}{2}+2\bigg[\frac{c_{\mathrm{d}}{}^{2}+c_{\mathrm{s}}{}^{2}}{2}+\frac{v_{0}{}^{2}\cos^{2}\alpha_{\mathbf{K}}}{4}\overset{\text{``\maths{1}''}}{\underset{\text{``\maths{2}''}}{\pm}}\sqrt{\frac{(c_{\mathrm{d}}{}^{2}-c_{\mathrm{s}}{}^{2})^{2}}{4}+\frac{(c_{\mathrm{d}}{}^{2}+c_{\mathrm{s}}{}^{2})v_{0}{}^{2}\cos^{2}\alpha_{\mathbf{K}}}{2}+\frac{v_{0}{}^{2}\cos^{2}\alpha_{\mathbf{K}}}{2}\frac{K^{2}}{2}}\bigg]\bigg\}}.
\end{equation}
The system is dynamically stable when both \maths{\Omega_{1}} and \maths{\Omega_{2}} are real-valued for all \maths{\mathbf{K}}. While \maths{\Omega_{1}} is so for any \maths{\mathbf{v}_{0}}, \maths{\Omega_{2}\in\mathbb{R}} for any \maths{\mathbf{K}} only when~\cite{Pitaevskii2016}
\begin{equation}
\label{Eq:DynStab}
v_{0}<2\min\{c_{\mathrm{d}},c_{\mathrm{s}}\}.
\end{equation}
At larger \maths{v_{0}}, a dynamical instability develops in the \maths{\mathbf{v}_{0}} direction. This is signaled by a nonzero support for \maths{\mathrm{Im}\,\Omega_{2}\neq0} at small wave numbers; as \maths{v_{0}} increases, this support detaches from the origin and shifts to larger wave numbers~\cite{Rodrigues2020}. In the dynamically stable regime~\eqref{Eq:DynStab}, \maths{\Omega_{1,2}} is of Bogoliubov type with sound velocity
\begin{equation}
\label{Eq:c12v}
c_{1,2}=\sqrt{\frac{c_{\mathrm{d}}{}^{2}+c_{\mathrm{s}}{}^{2}}{2}+\frac{v_{0}{}^{2}\cos^{2}\alpha_{\mathbf{K}}}{4}\overset{\text{``\maths{1}''}}{\underset{\text{``\maths{2}''}}{\pm}}\sqrt{\frac{(c_{\mathrm{d}}{}^{2}-c_{\mathrm{s}}{}^{2})^{2}}{4}+\frac{(c_{\mathrm{d}}{}^{2}+c_{\mathrm{s}}{}^{2})v_{0}{}^{2}\cos^{2}\alpha_{\mathbf{K}}}{2}}}.
\end{equation}
In the limit \maths{v_{0}\to0}, \maths{c_{1}\simeq\max\{c_{\mathrm{d}},c_{\mathrm{s}}\}} and \maths{c_{2}\simeq\min\{c_{\mathrm{d}},c_{\mathrm{s}}\}} obviously coincide with the density and spin speeds of sound characterizing the balanced configuration studied in the main text. Landau's critical speeds for energetic stability with respect to the ``\maths{1}'' and ``\maths{2}'' modes are given by
\begin{align}
\label{Eq:Landau1}
V_{1}&=\min_{\mathbf{K}}\frac{\Omega_{1}}{K}=c_{1}|_{\alpha_{\mathbf{K}}\in\{\pi/2,3\pi/2\}}=\max\{c_{\mathrm{d}},c_{\mathrm{s}}\}, \\
\label{Eq:Landau2}
V_{2}&=\min_{\mathbf{K}}\frac{\Omega_{2}}{K}=c_{2}|_{\alpha_{\mathbf{K}}\in\{0,\pi\}}=\sqrt{\frac{c_{\mathrm{d}}{}^{2}+c_{\mathrm{s}}{}^{2}}{2}+\frac{v_{0}{}^{2}}{4}-\sqrt{\frac{(c_{\mathrm{d}}{}^{2}-c_{\mathrm{s}}{}^{2})^{2}}{4}+\frac{(c_{\mathrm{d}}{}^{2}+c_{\mathrm{s}}{}^{2})v_{0}{}^{2}}{2}}},
\end{align}
and since \maths{V_{2}<V_{1}} for all \maths{v_{0}}, \maths{V_{2}} is Landau's critical speed \maths{V_{\mathrm{c}}} for superfluidity in this velocity-imbalanced configuration.

\subsection{\maths{\boldsymbol{\sigma_{0}\neq0}} and \maths{\boldsymbol{\mathbf{v}_{0}=0}}}
\label{SubApp:ImbalancedDensity}

Here, the magnetization angle at infinity \maths{\theta\simeq\theta_{0}\neq\pi/2}, the chemical potential \maths{\mu_{\pm}=(\mu_{0,\pm}+V_{0}{}^{2})/2} with \maths{\mu_{0,\pm}=\mu_{0}\{1\pm[(1-\alpha)/(1+\alpha)]\cos\theta_{0}\}}, and the matrix \maths{\mathbb{L}} reads
\begin{equation}
\label{Eq:Lsigma}
\mathbb{L}=
\begin{bmatrix}
0 & \frac{K^{2}}{2} & 0 & \frac{\cos\theta_{0}K^{2}}{2} \\
-\frac{K^{2}}{2}-2c_{\mathrm{d}}{}^{2} & 0 & -\frac{K^{2}}{2\tan\theta_{0}} & 0 \\
0 & 0 & 0 & -\frac{\sin\theta_{0}K^{2}}{2} \\
-2c_{\mathrm{d}}{}^{2}\frac{1-\alpha}{1+\alpha}\cos\theta_{0} & 0 & \frac{K^{2}}{2\sin\theta_{0}}+2c_{\mathrm{s}}{}^{2}\sin\theta_{0} & 0
\end{bmatrix}
.
\end{equation}
The total and relative linearized fluctuations of the superfluid are also hybridized in this case. Their dispersion relations are of Bogoliubov type, given by \maths{\Omega_{0}=\pm\Omega_{1}} and \maths{\Omega_{0}=\pm\Omega_{2}} with
\begin{align}
\label{Eq:Omega12sigma}
\Omega_{1,2}&=\sqrt{\frac{K^{2}}{2}\bigg(\frac{K^{2}}{2}+2c_{1,2}{}^{2}\bigg)}, \\
\label{Eq:c12sigma}
c_{1,2}&=\sqrt{\frac{c_{\mathrm{d}}{}^{2}}{2}\bigg(1+\frac{1-\alpha}{1+\alpha}\cos^{2}\theta_{0}\bigg)+\frac{c_{\mathrm{s}}{}^{2}}{2}\sin^{2}\theta_{0}\overset{\text{``\maths{1}''}}{\underset{\text{``\maths{2}''}}{\pm}}\sqrt{\bigg[\frac{c_{\mathrm{d}}{}^{2}}{2}\bigg(1+\frac{1-\alpha}{1+\alpha}\cos^{2}\theta_{0}\bigg)+\frac{c_{\mathrm{s}}{}^{2}}{2}\sin^{2}\theta_{0}\bigg]^{2}-c_{\mathrm{d}}{}^{2}c_{\mathrm{s}}{}^{2}\sin^{2}\theta_{0}}}.
\end{align}
In the limit \maths{\theta_{0}\to\pi/2}, we verify that \maths{c_{1}\simeq\max\{c_{\mathrm{d}},c_{\mathrm{s}}\}} and \maths{c_{2}\simeq\min\{c_{\mathrm{d}},c_{\mathrm{s}}\}} coincide with the density and spin speeds of sound of the fully balanced configuration. By construction, \maths{c_{1}} and \maths{c_{2}} are Landau's critical speeds for energetic stability of the flow with respect to the ``\maths{1}'' and ``\maths{2}'' modes, respectively, and since \maths{c_{2}<c_{1}} for all \maths{\theta_{0}}, \maths{c_{2}} is the critical speed \maths{V_{\mathrm{c}}} for dissipationless motion, within perturbation theory.

\end{strip}

\section{Numerical methods}
\label{App:Num}

A standard split-step Fourier method~\cite{Agrawal2019} is used to solve the coupled equations~\eqref{Eq:dNLS}. In this method, the complex field \maths{\psi_{\pm}(\mathbf{r},z)} is obtained through the following incremental relation:
\begin{equation}
\label{Eq:SplitStep}
\psi_{\pm}(\mathbf{r},z+\delta z)=\mathcal{F}^{-1}\Big\{e^{-i\frac{K^{2}}{2}\delta z}\mathcal{F}\Big[e^{-i\hat{P}_{\pm}\delta z}\psi_{\pm}(\mathbf{r},z)\Big]\Big\},
\end{equation}
where \maths{\mathcal{F}} denotes the space Fourier transform, \maths{K} is the magnitude of the transverse wave vector \maths{\mathbf{K}=K_{x}\hat{\mathbf{x}}+K_{y}\hat{\mathbf{y}}}, and \maths{\hat{P}_{\pm}=U_{\pm}(\mathbf{r})+(\rho_{\pm}+\alpha\rho_{\mp})/(1+\beta\rho)} describes the obstacle and interaction potentials for the \maths{\pm} wave. In the simulations, we use a smoothed version of the two-amplitude sharp disk~\eqref{Eq:U+-}, of the form
\begin{equation}
\label{Eq:UpmNum}
U_{\pm}(\mathbf{r})=\frac{U_{\pm}}{2}\bigg(1+\tanh\frac{w-r}{\delta w}\bigg),
\end{equation}
which is more amenable to our numerical scheme. We consider a radius \maths{w=10}, which is large enough to match the hydraulic regime studied theoretically, and a boundary width \maths{\delta w=1} to avoid numerical difficulties at short length scales. The initial light spots are top-hat-shaped with equal intensities, \maths{\rho_{\pm,0}=1/2}, and equal angles with respect to the \maths{z} axis, \maths{\mathbf{v}_{\pm,0}=V_{0}\hat{\mathbf{x}}}:
\begin{equation}
\label{Eq:InitialWaves}
\psi_{\pm}(\mathbf{r},z=0)=\frac{e^{iV_{0}x}}{\sqrt{2}}e^{-r^{6}/R^{6}}.
\end{equation}
The super-Gaussian weight \maths{\exp(-r^{6}/R^{6})} defines the radial extension of the two initial beams. In the simulations, we take \maths{R=140}. To minimize transient nonequilibrium features arising from the abrupt introduction of the obstacle and interaction potentials at the entrance, \maths{z=0} interface~\cite{Abuzarli2022, Scoquart2021}, both potentials are adiabatically ramped from zero to their maximum values over a few nonlinear lengths \maths{1/\mu=1}, and the propagation is computed until the system reaches a nearly stationary state.

\bibliographystyle{sn-aps}
\bibliography{2ComponentSF}

\end{document}